# Source localization realizes single frame super-resolution for fluorescence imaging

## Author Information

Mengrui Wang[1], Shouwen Ma[2], Zewei Luo[3], Wei Shi[4], Yiming Li[4], Yuwei Huang[5], Hu Zhao[6], Chang Liu[1], Manming Shu[1], Jingxiang Zhang[1], Yansheng Liang[1], Tianyu Zhao[1], Shaowei Wang[1], Tongsheng Chen[3], Chenguang Wang[7], Ming Lei[1,8*]

[1] MOE Key Laboratory for Nonequilibrium Synthesis and Modulation of Condensed Matter, School of Physics, Xi'an Jiaotong University, Xi'an, 710049, China

[2] Department Behavioural Neurobiology, Max Planck Institute for Biological Intelligence, Seewiesen, 82319, Germany

[3] MOE Key Laboratory of Laser Life Science & Institute of Laser Life Science, College of Biophotonics, School of Optoelectronic Science and Engineering, South China Normal University, Guangzhou, 510631, China

[4] Department of Biomedical Engineering, Southern University of Science and Technology, Shenzhen, 518055, China

[5] School of Basic Medical Sciences, Translational Medicine Institute, Xi'an Jiaotong University, Xi'an, 710049, China

[6] Chinese Institute for Brain Research, Zhongguancun Life Science Park, Beijing, 100086, China

[7] State Key Laboratory of Integrated Optoelectronics (JLU Region), College of Electronic Science & Engineering, Jilin University, Changchun, 130012, China

[8] State Key Laboratory of Electrical Insulation and Power Equipment, Xi'an Jiaotong University. Xi'an, 710049, China

\* Corresponding author: Ming Lei

Email: ming.lei@mail.xjtu.edu.cn

## Abstract

Existing super-resolution microscopy is often constrained by inherent trade-offs between resolution, acquisition speed, phototoxicity, and hardware complexity. Computational post-processing approaches offer a promising alternative, but they typically suffer from linearity distortion, high computational cost, reliance on pre-training data, or reconstruction artifacts. Here, we present Source Localization (SoLo), a novel single-frame super-resolution algorithm for fluorescence imaging without these limitations. Built on the principle of inferring fluorescent source positions via sampling-detection strategy, SoLo achieves non-iterative, parallelizable computation, enabling real-time live-cell imaging with high spatiotemporal resolution. The intensity linearity preservation of SoLo makes it compatible with quantitative analysis such as calcium imaging and fluorescence resonance energy transfer. We further extended this framework to 3D-SoLo for volumetric imaging and nonlinear SoLo (NL-SoLo) for high-density fluorescence fluctuation imaging. With its ease of parameter tuning and compatibility with existing imaging systems, SoLo offers an accessible solution for ordinary labs, enabling diverse biomedical imaging applications.

## Introduction

Fluorescence microscopy has long served as an indispensable tool in biomedical research, enabling the visualization of cellular and subcellular structures with molecular specificity[1,2]. However, the spatial resolution of conventional fluorescence microscopy is fundamentally constrained by the optical diffraction limit, which restricts the detailed observation of fine subcellular architectures and dynamic molecular interactions[3,4]. To overcome this barrier, various super-resolution (SR) methods have been developed, including structured illumination microscopy (SIM)[5], stimulated emission depletion microscopy (STED)[6], and single-molecule localization microscopy (SMLM)[7]. SIM uses patterned light to create moiré fringes that encode high-frequency information, offering the advantages of relatively low phototoxicity and high speed, but its resolution gain is limited to a factor of two and it is prone to reconstruction artifacts[8-10]. SMLM techniques pinpoint the center of individual fluorophores with nanometer precision by temporally separating their blinking, offering the highest spatial resolution, but at the cost of long acquisition times and intensive computational processing[11,12]. A variant of SMLM is fluorescence fluctuation (FF) imaging, such as super-resolution optical fluctuation imaging (SOFI)[13,14] and super-resolution radial fluctuation (SRRF)[15,16], which excites fluorophores at relatively high density and reconstructs images from temporal fluctuations rather than performing explicit single-molecule localization, greatly reducing the image acquisition requirements of SMLM.

The resolution gains provided by the SR methods described above are invariably accompanied by trade-offs, including increased image acquisition, higher phototoxicity, more complex systems, or specialized fluorescent probes. Alternatively, certain image post-processing approaches can also enhance the resolution of individual images without these penalties. One class of methods is based on iterative Richardson-Lucy (RL) deconvolution with regularization constraints—such as Sparse deconvolution incorporating sparsity constraints[17], MRA deconvolution embedding multi-resolution analysis[18], and 3Snet-CLID applying deep learning network denoising before deconvolution[19]. These approaches effectively suppress noise but suffer from high computational cost due to their iterative nature. Another class uses deep learning for resolution enhancement[20], such as the deep Fourier channel attention network (DFCAN)[21] and zero-shot deconvolution networks (ZS-DeconvNet)[22], which excels at producing high-resolution images close to the ground truth but rely on pre-training with known samples. There are also methods that improve resolution by sharpening the point spread function (PSF), including deblurring by pixel reassignment (DPR)[23], which reassigns pixels along gradient directions, and mean shift super-resolution (MSSR)[24], which reduces the PSF through neighborhood calculations. However, these methods often introduce strong nonlinearities or artifacts (although the image intensity is conserved in DPR, the local PSF becomes deformed due to varying degrees of modification).

We have proposed a source localization (SoLo) algorithm that achieves single-frame resolution enhancement for fluorescence images, offering comprehensive advantages such as fast computation, strong resolving power, high fidelity, and ease of use. It achieves high-resolution image restoration by inferring the positions of fluorescent sources and attains high computational efficiency through a sampling-detection strategy. For a 512×512 pixel image, it requires only 0.03 seconds of processing time, thereby enabling real-time SR and allowing users to quickly adjust parameters. The intensity linearity preservation of SoLo makes it suitable for quantitative analysis, such as calcium imaging[25] and fluorescence resonance energy transfer (FRET)[26]. Building on the advantage of simple computational model, we extended SoLo to three dimensions (3D-SoLo), which can simultaneously improve the relatively poor axial resolution in 3D imaging. It is applicable to various 3D imaging techniques, such as confocal microscopy[27], two-photon (2P) microscopy[28,29], optical sectioning structured illumination microscopy (OS-SIM)[30], and even wide-field (WF) microscopy under weak background. We demonstrated that WF imaging equipped with SoLo achieves high-resolution recording of rapid dynamics and long-term movements of living cells.

By introducing higher-order derivative terms of the image, we further developed the nonlinear SoLo (NL-SoLo). NL-SoLo possesses an ultra-high resolution capability for adjacent fluorophores, achieving an improvement of approximately 7-fold, while maintaining the advantage of high-speed computation. When applied to SMLM, NL-SoLo allows for higher-density fluorophore activation, thereby significantly enhancing its imaging efficiency. In high-density FF imaging, NL-SoLo surpasses single-molecule fitting method and eSRRF in terms of reconstruction resolution, computational speed, and intensity linearity.

## Results

### Principle and performance of SoLo

SoLo is based on the assumption that the point spread function (PSF) follows a Gaussian-like distribution. For Gaussian distribution $I(x) = A_0 e^{-\|x-x_0\|^2/2\sigma_{PSF}^2}$, the $\sigma_{PSF}^2 \nabla I(x)/I(x)$ at any position $x$ is directed toward its central point $x_0$ (Supplementary Notes 1). This provides an approach to infer the location of fluorescent sources by analyzing the intensity and gradient of the image. Directly reassigning all pixels along their gradients to the inferred source locations could enhance resolution, as implemented in DPR[23]. However, this implementation involves frequent scattering and gathering operations, resulting in low processing efficiency, and primarily improves edge resolution while offering limited enhancement for internal structures.

SoLo employs a sampling detection strategy to determine the probability of each location being an emission source in the fluorescence image $I$ (Fig.1a). For the target detection position $x$, we uniformly sample $N$ points $s_i$ around with radius $r$ ($N = 8$ for 2D image; Supplementary Notes 2, Supplementary Fig.2a) and calculate their potential source locations using

$$D(x) = \sigma_{PSF}^2 \nabla I(x) / I(x). \tag{1}$$

We model the true source position as following a Gaussian probability distribution centered at the located source. Thus, the detection position receives radiation from the located source according to a Gaussian function with distance-dependent attenuation and variance $\sigma_r$. The final SoLo image is calculated as

$$I_{\text{SoLo}}(\boldsymbol{x}) = \frac{1}{N} \sum_{i=1}^{N} C e^{-\|\boldsymbol{s}_i + \boldsymbol{D}(\boldsymbol{x}+\boldsymbol{s}_i)\|^2 / 2\sigma_r^2} I(\boldsymbol{x}+\boldsymbol{s}_i), \qquad (2)$$

where $C$ serves as a normalization coefficient to maintain intensity conservation between $I_{\text{SoLo}}$ and $I$. For images with low sampling rates, it is advisable to perform upsampling in preprocessing. Given the sensitivity of gradient computation, the upsampling should be executed in the frequency domain, coupled with denoising according to the cutoff frequency (Supplementary Notes 3).

The performance of SoLo algorithm is primarily influenced by 4 parameters: full width at half maximum (FWHM) of PSF *psf* (and then $\sigma_{PSF} = psf / 2\sqrt{2\ln 2}$), upsampling factor, sampling radius *r*, and $k = \sigma_{PSF}/\sigma_r$ as PSF shrinking factor. Determining the optimal parameter ranges and understanding their impact on SoLo algorithm performance can significantly facilitate parameter tuning for optimal results. *psf* can be calculated based on the theoretical diffraction limit determined by the objective's numerical aperture (NA) and emission wavelength. Alternatively, it may be derived from image using resolution measurement methods[31, 32], or even estimated directly from fine structural details. To achieve more uniform radius of 8-point sampling, the image sampling rate should preferably be increased to at least 5/*psf* (Supplementary Notes 2, Supplementary Fig.2a). Through dot-array simulation, we investigated the effects of *r* and *k* on the resolution and fidelity of output images (Supplementary Notes 4, Extended Data Fig.1). Based on the comprehensive metric "Quality and Resolution" (QnR), the optimal performance was achieved when $r \approx 0.3psf$ and $k \approx 4$, with the constraint that $r \cdot k < 2.6psf$. We further demonstrated the effects of varying *psf*, *r*, and *k* through both line-resolution simulation and experimental mitochondrial imaging (Supplementary Notes 5, Supplementary Fig.3). The results revealed that using a *psf* slightly smaller than the actual value enhances resolution, but an excessively small *psf* introduces artifacts. Besides, smaller *r* and *k* values result in lower resolution, while overly large values cause hollowing-out effects in the sample's interior structure.

To evaluate SoLo's resolution enhancement, we performed standard resolution measurement using Argo-SIM slide containing fluorescent line pairs with varying spacings (Fig.1b). Under WF imaging mode, only line pairs ≥210nm were resolvable. SoLo improved this to 120nm, matching the performance of SR-SIM, and further enhanced SIM images to 60nm. These results demonstrate that SoLo can achieve approximately 2-fold resolution improvement in sparse condition.

We next benchmarked SoLo against MSSR, DPR, and Sparse deconvolution using biological fluorescence images from the BioSR dataset[21] and BF-SIM[33], with SIM images as ground truth (GT) (Fig.1c). Three representative structures were examined: endoplasmic reticulum (ER), clathrin-coated pits (CCPs), and mitochondria. For ER which appears a sparse, mesh-like structure: MSSR exhibited nonlinear intensity loss; DPR showed improved resolution but with artifacts; Sparse deconvolution enhanced image signal-to-noise ratio (SNR) and contrast but provided mere resolution improvement; while SoLo achieved resolution enhancement closest to SIM. CCPs exhibits a central depression with a ring-like structure, but WF imaging lacks the resolution to resolve this hollow morphology, rendering them as solid puncta. While MSSR and DPR merely reduced spot size without revealing internal structure, Sparse deconvolution and SoLo successfully resolved the hollow morphology. Mitochondrial cristae represent densely packed structures that appear as blurred, continuous intensity distributions in WF image, making them among the most challenging targets for computational resolution enhancement. MSSR severely distorted mitochondrial morphology, DPR introduced extensive internal artifacts, and Sparse deconvolution provided negligible improvement. Remarkably, SoLo retained the ability to partially resolve cristae structures under these conditions. Quantitative evaluation using PSNR (with SIM as reference) and fidelity (with WF) confirmed that SoLo consistently achieved the best balance of resolution and structural fidelity across all three specimens (Fig.1d).

Due to computationally simple, non-iterative, and parallelizable architecture, SoLo offers flying processing speed. We measured the processing speed of each algorithm on a system with AMD Ryzen 5 3600X CPU and NVIDIA GeForce RTX 3070 GPU (Fig.1e). While Sparse deconvolution's iterative optimization resulted in very slow processing, SoLo achieved the fastest performance - requiring just 0.08s for 1024×1024 pixels and 0.03s for 512×512 pixels, representing a >200× speed advantage over Sparse deconvolution. This rapid processing enables real-time resolution enhancement. Through a dual-thread acquisition-processing pipeline, we demonstrated live super-resolution up to 30fps @ 512×512 pixels, supporting in situ observation of rapid biological processes (Supplementary Notes 6, Supplementary Fig.4 and Supplementary Video 1). Besides, parameter adjustments also provide instant visual feedback, allowing users to optimize settings without processing delays (Supplementary Video 2).

## SoLo's single-frame super-resolution is beneficial for live-cell dynamics imaging

In live-cell imaging, multiple factors such as spatial resolution, temporal resolution, and light dose are often in conflict. For example, SIM enhances spatial resolution by acquiring multiple frames, but at the cost of reduced temporal resolution and increased light exposure. Capturing rapid interactions among organelles requires sufficiently high temporal resolution, while observing long-term cellular movements and changes demands a low light dose to minimize phototoxicity and photobleaching. SoLo achieves super-resolution from a single frame, making it advantageous for high-speed or long-term live-cell imaging.

Migrasomes are organelles generated during cell migration, functioning in materials and information transfer as well as in establishing micro-environments. We recorded the retrieval process of migrasomes in NRK cells under room temperature using WF imaging, followed by SoLo enhancement (Fig. 2a–d, Supplementary Video 3). Over 172 minutes, migrasomes transported retrogradely along retraction fibers toward the cell body, accompanied by a progressive decrease in size. When their diameter reduced below 500 nm, WF imaging could no longer resolve the membrane, whereas SoLo extended the resolvable limit down to 200 nm, enabling clearer visualization of the migrasome retrieval process (Fig. 2b). Some migrasomes were retrieved via migracytosis-like mechanism: they first elongated along the retraction fiber and were then severed in the middle—one portion of the extruded migrasome was transported toward the cell, while the remaining portion stayed in place (Fig. 2c). During retrieval, the retraction fibers themselves also underwent morphological changes, attempting to form new, shorter, and straighter fibers at the concave sides of certain curvatures (Fig. 2d). These results underscore the advantage of SoLo in resolving fine structures below the diffraction limit. Furthermore, the observed retrieval process raises the possibility that migrasomes may participate in the retrograde transport of materials and signals from the cell periphery back to the cell body, a previously underexplored direction in migrasome biology.

Real-time SR imaging enables direct, instant, and interactive high-resolution observation during live cell dynamic processes, avoiding the limitation of missing key transient events due to post-acquisition data processing. We applied SoLo to real-time SR imaging of COS-7 cell mitochondria (Fig. 2e-j, Supplementary Video 1). The cristae structures of mitochondria, which were blurred in WF images, became clearly visible after enhancement by SoLo (Fig. 2f,g), allowing temporal observation of cristae remodeling. Multiple events such as mitochondrial extension and kissing, rotation of ring- or circle-shaped mitochondria, as well as fission and fusion, were observed (Fig. 2h-j).

Because the depth of field of a high-NA objective often cannot cover the entire cell thickness, leading to the loss of out-of-focus information, we performed multi-layer scanning of COS-7 cell mitochondria using WF imaging (Fig. 2k-o, Supplementary Video 3). SoLo was applied to enhance the resolution of each layer, followed by background removal using the rolling ball method for stacked MIP generation. Cristae structures in different layers, indistinct in WF, became clearly visible after SoLo (Fig. 2l), and inner membrane changes during fusion were resolved (Fig. 2m). Notably, we observed that the mitochondria in this pair of newly divided cells exhibited a filamentous interconnected network, with many mitochondria still showing extension behaviors and attempting to dock with adjacent mitochondria (Fig. 2n), and the established mitochondrial network continuously underwent remodeling (Fig. 2o). We speculate that these transiently established channels may allow the exchange of contents between mitochondria, facilitating local quality control and energy transfer.

The dynamics of ER is a relatively rapid process. We utilized ER data[34] acquired by WF imaging to demonstrate the capability of SoLo in resolving fast dynamic behaviors (Extended Data Fig. 2, Supplementary Video 3). SoLo resolved fine mesh structures within the ER network and distinguished adjacent tubules while suppressing background. Morphological changes of the ER, including extension, contraction, and fusion, were observed with greater clarity and accuracy.

SoLo enables quantitative analysis of fluorescence intensity

Super-resolution methods often suffer from intensity linearity distortion, making them inapplicable for quantitative analysis. A typical example is SIM, whose image reconstruction focuses on the frequency spectrum and fails to maintain the linear correlation between intensity and fluorescence signals. Moreover, it is prone to reconstruction artifacts, compromising the accuracy of quantitative analysis. Image optimization algorithms based on deconvolution typically impose nonlinear constraints or prior knowledges, which lead to the disruption of intensity linearity.

As a non-iterative spatial-domain method, SoLo computes from neighboring pixel intensities, offering high fidelity and preserving intensity linearity. We applied SoLo to calcium imaging in freely moving mice using miniaturized two-photon microscope (MINI2P)[35], where calcium concentration correlates with fluorescence intensity to reflect neuronal activity (Supplementary Video 4). Denoising and resolution enhancement by SoLo revealed finer morphological details with improved clarity and contrast (Extended Data Fig. 3a,b). Measurements showed that the resolution improved from 6.44 μm to 3.77 μm, and the SNR increased from 5.07 dB to 15.86 dB (Extended Data Fig. 3d,e). This enabled detection of more neurons: Cellpose segmentation[36] identified 91 neurons from raw 2P images but 166 after SoLo (Extended Data Fig. 3c,f). A heatmap of calcium signals from all neurons detected by SoLo was generated and clustered using k-means (Extended Data Fig. 3g). For the six neurons marked in Extended Data Fig. 3a, raw and SoLo-processed signals were nearly identical (Extended Data Fig. 3h), demonstrating the excellent intensity linearity fidelity of SoLo.

FRET imaging is highly sensitive to the linearity of fluorescence intensity, as even minor reconstruction artifacts can lead to significant errors in calculating FRET efficiency ($E_D$) and the acceptor-to-donor concentration ratio ($R_C$). To evaluate the intensity fidelity and the practical utility of SoLo in sub-diffraction functional imaging, we performed quantitative FRET analysis in live U2OS cells expressing the mitochondria-targeted calibration plasmid ACTA-G17M (theoretical $E_D = 0.2$, $R_C = 1$). We assessed the grayscale linear fidelity by comparing the intensities of the donor-donor (DD), donor-acceptor (DA), and acceptor-acceptor (AA) channels against conventional WF images (Fig. 3a). While SIM reconstruction introduces intensity deviations and artifacts due to complex modulation patterns, SoLo maintained an excellent linear correlation with WF intensities ($R^2 = 0.96$), demonstrating superior quantitative preservation without the need for external hardware modulation (Fig. 3b). Furthermore, SoLo's resolution enhancement allowed for more precise spatial mapping of FRET metrics on mitochondrial membranes compared to diffraction-limited WF imaging (Fig. 3c,e). Line profiles confirmed that SoLo achieves a spatial resolution comparable to SIM, effectively breaking the diffraction limit with enhanced peak-to-background contrast (Fig. 3g,h). Investigation into the distributional characteristics of FRET metrics indicates that while SIM-FRET is inherently quantitative, its reconstruction process—susceptible to structure-dependent noise and frequency-domain artifacts—leads to a characteristic broadening of $E_D$ and $R_C$ histograms[37] (Fig. 3d,f). Notably, SoLo-enhanced images exhibit a quantitative fidelity that aligns with the WF "gold standard", suppressing the reconstruction-induced fluctuations inherent in SIM while preserving narrow and accurate distributions. Statistical evaluation across multiple fields of view (Fig. 3i,k) further validates that SoLo resolves fine structural details without compromising quantitative precision.

## 3D-SoLo enhances both lateral and axial resolution in 3D imaging

Three-dimensional imaging technology can obtain complete structural information of a sample, and its axial resolution directly determines the ability to distinguish details of adjacent layers in the depth direction. To enhance both the lateral and axial resolution of three-dimensional imaging, we extended SoLo to 3D mode. It is based on the assumption that the 3D PSF follows an ellipsoidal Gaussian distribution. The implementation principle of 3D-SoLo remains unchanged, with the main difference being that the sampling point distribution shifts from a circular ring to an ellipsoidal surface—employing different sampling radii $r_{xy}$ and $r_z$ in the lateral and axial directions according to $psf_{xy}$ and $psf_z$. To ensure both high efficiency and uniformity of sampling detection, we adopted a sampling strategy with 14 points (Supplementary Fig. 5). Correspondingly, the PSF shrinking factor $k$ can also be set separately for $k_{xy}$ and $k_z$ to emphasize resolution enhancement in specific directions—e.g., specifically reducing the axial PSF to achieve near-isotropic imaging for better 3D morphological analysis. Using 3D-SIM as GT, we validated that 3D-SoLo achieves a 1.57-fold lateral and 2-fold axial resolution enhancement, producing results closely resembling those of 3D-SIM (Supplementary Notes 7, Extended Data Fig. 4).

3D neural imaging is crucial for revealing the complex structural architecture and dynamic functional networks of the brain, providing essential insights that bridge molecular mechanisms with systemic understanding in fundamental neuroscience. We performed 2P imaging of neurons in the HVC region of the zebra finch followed by 3D-SoLo enhancement. The enhanced MIP images revealed neural structures with improved clarity and higher SNR, restoring originally fragmented neural fibers into continuous morphology (Fig. 4a,b). Axial resolution was improved from 5.8 ± 1.3 μm to 2.9 ± 0.6 μm, approximately doubled, enabling clear 3D visualization of neural structures (Fig. 4c, Supplementary Video 5). As imaging depth increased, the SNR of 2P imaging gradually decreased by 9.3 dB, whereas 3D-SoLo maintained it at a high level, with 7.3 dB higher in deep regions (Fig. 4d). 3D-SoLo also enabled the detection of a greater number of neurons, increasing from 49 to 84 (Fig. 4e), and clearly revealed the changes in spines and boutons on different days (Fig. 4f). Detailed comparisons of deep-layer regions illustrated the ability of 3D-SoLo to enhance neural fibers that were originally difficult to visualize, facilitating more accurate analysis of their morphological changes.

We also demonstrated the enhancement of confocal imaging by 3D-SoLo using section of *Thy1-EGFP-M* mouse brain as sample. 3D-SoLo improved both lateral and axial resolution, as well as the SNR, with axial resolution approximately doubled (Extended Data Fig. 5, Supplementary Video 5). Detailed structural features of the neurons were enhanced, and densely distributed dendritic spines became clearly visible.

We further demonstrated the improvement of 3D-SoLo for live-cell 3D imaging by recording the 3D dynamics of lipid droplets over 10 minutes using WF microscopy (Extended Data Fig. 6, Supplementary Video 5). In 3D imaging, WF suffers from strong background and poor axial PSF. 3D-SoLo substantially mitigated these issues, greatly enhancing contrast and improving the lateral FWHM of lipid droplet PSF from 454 nm to 319 nm, and the axial FWHM from 1155 nm to up to 322 nm, approaching isotropy. This enabled better resolution of the 3D dynamic movements of lipid droplets, including processes such as fusion and fission.

## NL-SoLo for fluorescence fluctuation-based super-resolution microscopy

By incorporating higher-order derivative relations, SoLo can be extended to form nonlinear-SoLo (NL-SoLo). The standard SoLo approach assumes intensity distributions from isolated point sources, which limits its sensitivity in regions between adjacent sources. If the image $I$ is modeled as the intensity distribution from two point sources, incorporating up to third-order derivatives enables precise localization of two closely spaced point sources—but at the cost of excessive computational burden. Through simplifying solving conditions, we achieve an approximate solution requiring only second-order derivatives with significantly reduced computation (Supplementary Notes 8). The solution augments the original gradient vector with separation vectors $\boldsymbol{S}_{1,2}$ to locate two potential point source positions:

$$\boldsymbol{D}_{1,2} = \sigma_{PSF}^2 \frac{\nabla I}{I} + \boldsymbol{S}_{1,2}, \tag{3}$$

$$\boldsymbol{S}_1 = -\boldsymbol{S}_2 = \sigma_{PSF} \left( \sqrt{\sigma_{PSF}^2 \frac{I \nabla_{xx}^2 I - (\nabla_x I)^2}{I^2} + 1}, \operatorname{sgn}(I \nabla_{xy}^2 I - \nabla_x I \nabla_y I) \sqrt{\sigma_{PSF}^2 \frac{I \nabla_{yy}^2 I - (\nabla_y I)^2}{I^2} + 1} \right), \tag{4}$$

where sgn() represents sign function. Consequently, NL-SoLo can detect both adjacent sources even in regions between them. This eliminates the need for radius-based sampling around detection locations—as required in standard SoLo—and enables direct source localization (Fig. 5a). The NL-SoLo image is generated similarly using a Gaussian-attenuation radiation received from the located sources:

$$I_{\text{NL-SoLo}} = \frac{C}{2} \left( e^{-\|\boldsymbol{D}_1\|^2 / 2\sigma_r^2} + e^{-\|\boldsymbol{D}_2\|^2 / 2\sigma_r^2} \right) I, \tag{5}$$

We employed simulated imaging data of selectively activated fluorescent molecules[38] to demonstrate the super-resolution capability of NL-SoLo. Multiple pairs of fluorescent molecules with varying inter-particle distances were selected from the simulated data, and their raw WF images were processed using eSRRF[16], multi-emitter maximum likelihood estimation (MLE) fitting[39], and NL-SoLo for resolution comparison (Fig. 5b). While WF microscopy could only resolve molecular pairs separated by 340.7 nm, both eSRRF and MLE fitting extended the resolution limit to 262.5 nm—yet failed at 157.4 nm separation. Remarkably, NL-SoLo achieved accurate resolution even for molecular pairs with a spacing of 48.7 nm, representing an approximately 7-fold resolution improvement compared to WF.

However, despite its enhanced resolution capability, NL-SoLo produces discretized images, which contradicts the continuous nature of most biological structures. Nevertheless, in single-molecule localization microscopy (SMLM) or fluorescence fluctuation (FF) imaging, fluorescent molecules are stochastically and discretely activated, making NL-SoLo applicable. For each raw image, the processing pipeline involves: upsampling to the target resolution, computing derivatives, solving for source localization and generating the NL-SoLo image (Fig. 5a). Unlike other FF reconstruction methods such as SRRF and MSSR, NL-SoLo does not require high-order temporal correlation (e.g., SOFI) to reduce FWHM: resolution is directly determined by the user-defined $σ_r$, thus requiring only mean temporal analysis—which additionally preserves intensity linearity.

To evaluate the localization accuracy of NL-SoLo, we employed synthetic data imitating microtubule structures[38]. The image formation model accounted for the stochastic nature of fluorophore emission, the characteristics of the optical setup, and various noise sources to generate data close to biological reality, which has been previously used for benchmarking various SMLM methods. The raw data consisted of 12,000 frames of low-density activated fluorophores with an average nearest-neighbor distance $<d_{NN}>$ of 1,133 nm. For comparison, we selected eSRRF as a representative non-explicit image reconstruction method—noted for its high-density applicability and high processing speed—alongside the multi-emitter MLE fitting provided by ThunderSTORM[39] as a representative explicit localization approach, which has been demonstrated to achieve localization precision closest to the theoretical Cramér-Rao lower bound (CRLB)[40]. The low-density reconstruction results show that both MLE fitting and NL-SoLo achieved superior resolution compared to eSRRF (Fig. 5c). We simultaneously measured the localization precision of each method (since eSRRF and NL-SoLo do not perform explicit localization, their positions were determined by local maxima) and plotted the relationship between precision and nearest-neighbor separation (Fig. 5d). We observed that both eSRRF and MLE fitting exhibited increased localization inaccuracy at separations below 300 nm and failed to resolve adjacent emitters completely at separations below 200 nm (where Precision = Separation/2 indicates the inability to resolve two adjacent fluorophores). In contrast, NL-SoLo's localization precision did not exhibit significant inaccuracy as the separation decreased, maintaining errors within 30 nm while achieving resolvable separation close to 50 nm. Although its localization accuracy for isolated fluorophores (separation >600 nm) did not surpass the theoretically optimal MLE fitting, it slightly outperformed eSRRF.

The high-precision resolution capability of NL-SoLo for closely spaced emitters indicates its significant advantage in reconstructing high-density fluorescence activation data. Accordingly, we performed 10-frame binning on the original data to simulate high-density activation, resulting in a reduced $<d_{NN}>$ of 260 nm, below the diffraction limit. eSRRF failed to resolve adjacent structures and exhibited intensity inhomogeneity; MLE fitting produced numerous erroneous localizations; while NL-SoLo maintained reconstruction quality closest to the GT (Fig. 5c). Statistical analysis of localization precision revealed that while MLE fitting performed optimally at low density, it significantly underperformed eSRRF and NL-SoLo at high density, with NL-SoLo consistently slightly outperforming eSRRF (Fig. 5d).

## NL-SoLo enables high-precision and rapid FF reconstruction

We employed nuclear pore complex (NPC) as a standard sample, given its known stereotypical structure, to evaluate the precision of NL-SoLo in practical FF imaging. Nup96 was identified as a suitable reference protein, forming a symmetrical eight-corner structure within each NPC[41]. FF recordings of Nup96-SNAP labeled with BG-AF647 were performed under low-density excitation for 100,000 frames. NL-SoLo reconstruction (upsampling factor = 20) successfully resolved the eight-corner structure of NPC (Fig. 6a,b). Compared against MLE fitting as the benchmark—which offers the highest accuracy under low-density excitation—NL-SoLo yielded highly similar results (Fig. 6c,d). Decorrelation analysis showed that NL-SoLo achieves a resolution of 14.6 nm, representing a 23.4-fold improvement over the raw WF and approaching the 12.6 nm resolution of MLE (Fig. 6e). This demonstrates that NL-SoLo, as a non-explicit localization FF reconstruction method, can achieve high precision comparable to explicit localization methods.

The limitation of explicit localization methods lies in their exclusive suitability for low-density excitation, resulting in inefficient imaging. NL-SoLo permits high-density excitation and offers the advantage of resolving closely adjacent fluorophores. We performed SR reconstruction on high-density FF imaging data of microtubules[42] using eSRRF, multi-emitter MLE fitting, and NL-SoLo, respectively (Fig. 6f,g). NL-SoLo achieved the best resolution, distinguishing three adjacent microtubule structures, while eSRRF resolved only two (Fig. 6h). Notably, NL-SoLo exhibited the best continuity in microtubule structure, while the MLE result appeared fragmented, probably due to its missed localizations of dense fluorophores. Intensity profiles along a segment of microtubule revealed that NL-SoLo maintained intensity presence across the entire segment, whereas MLE lost intensity at multiple points, and eSRRF fell between the two (Fig. 6i). This indicates that NL-SoLo offers better intensity linearity and is more effective at preserving the integrity of biological structures. Decorrelation analysis revealed that NL-SoLo achieved the best resolution of 46 nm, representing a 7.7-fold improvement over WF imaging (Fig. 6j). Most importantly, the explicit localization of high-density fluorophores by MLE was highly inefficient in terms of processing time, requiring 2 days, 10 hours, and 24 minutes to complete—an obviously impractical duration (Fig. 6k). In contrast, the processing time for NL-SoLo was only 4.6 minutes, representing a remarkable 763-fold increase in reconstruction efficiency.

## Discussion

The development of SR imaging and computational methods has advanced our understanding of subcellular processes, yet the practical deployment remains constrained by limitations in resolution, speed, and fidelity. In this work, we introduced the SoLo algorithm, a robust, single-frame resolution enhancement approach that can be seamlessly integrated with existing imaging systems. Leveraging the advantages of SoLo in high resolution, fast processing, and high-fidelity reconstruction, it overcomes the key limitations of existing technologies across diverse application scenarios.

WF microscopy offers high temporal resolution, low phototoxicity and simple hardware for live-cell dynamic imaging, but its spatial resolution is diffraction-limited. SoLo enhances WF resolution without hardware modification or additional phototoxicity, enabling simultaneous high spatiotemporal resolution for tracking rapid dynamic, such as migrasome transport and ER tubule remodeling. More critically, the high computational speed of SoLo enables real-time processing and visualization from WF to SR imaging, which is particularly valuable for discovering novel fast, transient biological processes.

Numerous SR techniques focus solely on resolution improvement while neglecting the preservation of intensity linearity, which severely hinders their application in quantitative analysis. This is particularly prominent in FRET assays, where the readouts are calculated via subtraction and division of fluorescence signals, rendering them extremely sensitive to nonlinear intensity distortions and reconstruction artifacts[43]. The intensity linearity preservation of SoLo enables reliable resolution enhancement for quantitative imaging modalities including calcium imaging and FRET, without compromising the linear relationship between signal and molecular activity.

Conventional SR methods, including SIM, STED, and SMLM, are typically limited to thin samples or single cells due to their limited penetration depth, as they rely on visible light illumination that suffers from severe scattering and absorption in thick tissues. In contrast, 2P microscopy, which uses near-infrared excitation, has become the gold standard for deep tissue imaging, but its spatial resolution is still constrained by the diffraction limit. By integrating 3D-SoLo with 2P, we can significantly enhance the spatial resolution of volumetric images acquired from deep brain tissue, generating high-resolution 3D data with a penetration depth far beyond traditional SR technique. This enables the study of subcellular neuronal structures in intact brain tissue, which is critical for studying large-scale neural circuit dynamics in vivo.

Traditional SMLM techniques rely on fluorophore separation for high-precision localization, requiring large number of frames over long period—limiting temporal resolution and increasing phototoxicity. NL-SoLo, with its ultra-high resolution capability, can resolve adjacent fluorophores even at high activation densities, thus requiring far fewer raw frames for SR reconstruction. This significantly reduces the acquisition time of single-molecule or FF imaging, making it possible to capture faster dynamic processes while reducing photobleaching and phototoxicity.

In summary, SoLo is a versatile computational tool that is compatible with a wide range of imaging modalities. It addresses the key limitations in quantitative analysis, deep tissue imaging, live-cell dynamics, and efficient single-molecule imaging. Its simple parameter tuning with real-time feedback and compatibility with existing imaging hardware allow ordinary biological laboratories to adopt SoLo without expensive upgrades. We believe that SoLo has the potential to become a routine tool for fluorescence imaging, democratizing SR imaging and enabling new discoveries in diverse areas of biomedical research.

## References

1. Tsien, R. Y. The green fluorescent protein. *Annu. Rev. Biochem.* **67**, 509-544 (1998). doi:10.1146/annurev.biochem.67.1.509.

2. Huang, B., Bates, M. & Zhuang, X. Super-resolution fluorescence microscopy. *Annu. Rev. Biochem.* **78**, 993-1016 (2009). doi:10.1146/annurev.biochem.77.061906.092014.

3. Schermelleh, L. et al. Super-resolution microscopy demystified. *Nat. Cell. Biol.* **21**, 72–84 (2019). doi:10.1038/s41556-018-0251-8

4. Hell, S. W. Far-Field Optical Nanoscopy. *Science* **316**, 1153-1158 (2007). doi:10.1126/science.1137395

5. Gustafsson, M. G. L. Surpassing the lateral resolution limit by a factor of two using structured illumination microscopy. *J. Microsc.* **198**, 82-87 (2000). doi:10.1046/j.1365-2818.2000.00710.x

6. Hell, S. W. & Wichmann, J. Breaking the diffraction resolution limit by stimulated emission: stimulated-emission-depletion fluorescence microscopy. *Opt. Lett.* **19**, 780-782 (1994). doi: 10.1364/ol.19.000780.

7. Betzig, E. et al. Imaging Intracellular Fluorescent Proteins at Nanometer Resolution. *Science* **313**, 1642-1645 (2006). doi:10.1126/science.1127344

8. Huang, X. et al. Fast, long-term, super-resolution imaging with Hessian structured illumination microscopy. *Nat. Biotechnol.* **36**, 451–459 (2018). doi:10.1038/nbt.4115

9. Zhao, T. et al. Real-time super-resolution structured illumination microscopy: current progress in joint space and frequency reconstruction. *Rep. Prog. Phys.* **88**, 076401 (2025). doi:10.1088/1361-6633/adecb1

10. Wang, Z. et al. Rapid, artifact-reduced, image reconstruction for super-resolution structured illumination microscopy. *Innovation* **4**, 100425 (2023). doi:10.1016/j.xinn.2023.100425

11. Rust, M., Bates, M. & Zhuang, X. Sub-diffraction-limit imaging by stochastic optical reconstruction microscopy (STORM). *Nat. Methods* **3**, 793–796 (2006). doi:10.1038/nmeth929

12. Lelek, M. et al. Single-molecule localization microscopy. *Nat. Rev. Methods Primers* **1**, 39 (2021). doi:10.1038/s43586-021-00038-x

13. Dertinger, T., Colyer, R., Iyer, G., Weiss, S., & Enderlein, J. Fast, background-free, 3D super-resolution optical fluctuation imaging (SOFI). *Proc. Natl. Acad. Sci. U.S.A.* **106**, 22287-22292 (2009). doi:10.1073/pnas.0907866106

14. Basak, S. et al. Super-resolution optical fluctuation imaging. *Nat. Photon.* **19**, 229–237 (2025). doi:10.1038/s41566-024-01571-3

15. Gustafsson, N. et al. Fast live-cell conventional fluorophore nanoscopy with ImageJ through super-resolution radial fluctuations. *Nat. Commun.* **7**, 12471 (2016). doi:10.1038/ncomms12471

16. Laine, R. F. et al. High-fidelity 3D live-cell nanoscopy through data-driven enhanced super-resolution radial fluctuation. *Nat. Methods* **20**, 1949–1956 (2023). doi:10.1038/s41592-023-02057-w

17. Zhao, W. et al. Sparse deconvolution improves the resolution of live-cell super-resolution fluorescence microscopy. *Nat. Biotechnol.* **40**, 606–617 (2022). doi:10.1038/s41587-021-01092-2

18. Hou, Y. et al. Multi-resolution analysis enables fidelity-ensured deconvolution for fluorescence microscopy. *eLight* **4**, 14 (2024). doi:10.1186/s43593-024-00073-7

19. Xue, F. et al. High-fidelity single-frame computational super-resolution using signal-preserving denoising-enabled deconvolution. *Nat. Commun.* (2026). doi:10.1038/s41467-026-70791-8

20. Li, S. et al. AI-empowered super-resolution microscopy: a revolution in nanoscale cellular imaging. *Nat. Methods* (2025). doi:10.1038/s41592-025-02871-4

21. Qiao, C. et al. Evaluation and development of deep neural networks for image super-resolution in optical microscopy. *Nat. Methods* **18**, 194–202 (2021). doi:10.1038/s41592-020-01048-5

22. Qiao, C. et al. Zero-shot learning enables instant denoising and super-resolution in optical fluorescence microscopy. *Nat. Commun.* **15**, 4180 (2024). doi:10.1038/s41467-024-48575-9

23. Zhao, B. & Mertz, J. Resolution enhancement with deblurring by pixel reassignment. *Adv. Photon.* **5**, 066004 (2023). doi:10.1117/1.AP.5.6.066004

24. Torres-García, E. et al. Extending resolution within a single imaging frame. *Nat. Commun.* **13**, 7452 (2022). doi:10.1038/s41467-022-34693-9

25. Zhong, J. et al. FACED 2.0 enables large-scale voltage and calcium imaging in vivo. *Nat. Methods* (2025). doi:10.1038/s41592-025-02925-7

26. Nettels, D. et al. Single-molecule FRET for probing nanoscale biomolecular dynamics. *Nat. Rev. Phys.* **6**, 587–605 (2024). doi:10.1038/s42254-024-00748-7

27. Qi, X. et al. Confocal Airy beam oblique light-sheet tomography for brain-wide cell type distribution and morphology. *Nat. Methods* **22**, 2622–2630 (2025). doi:10.1038/s41592-025-02888-9

28. Zhao Z. et al. Two-photon synthetic aperture microscopy for minimally invasive fast 3D imaging of native subcellular behaviors in deep tissue. *Cell* **186**, 2475-2491.e22 (2023). doi: 10.1016/j.cell.2023.04.016.

29. Qian, et al. High-throughput two-photon volumetric brain imaging in freely moving mice. *Nat. Commun.* **17**, 206 (2026). doi:10.1038/s41467-025-66922-2

30. Wang, M. et al. Structured illumination microscopy for high SNR 3D imaging from millimeter-thick tissues to cellular dynamics. *Innovation* (2026). doi:10.1016/j.xinn.2026.101321

31. Descloux, A., Grußmayer, K. S. & Radenovic, A. Parameter-free image resolution estimation based on decorrelation analysis. *Nat. Methods* **16**, 918–924 (2019). doi:10.1038/s41592-019-0515-7

32. Koho, S. et al. Fourier ring correlation simplifies image restoration in fluorescence microscopy. *Nat. Commun.* **10**, 3103 (2019). doi:10.1038/s41467-019-11024-z

33. Mo, Y. et al. Quantitative structured illumination microscopy via a physical model-based background filtering algorithm reveals actin dynamics. *Nat. Commun.* **14**, 3089 (2023). doi:10.1038/s41467-023-38808-8

34. Cao, R. et al. Dark-based optical sectioning assists background removal in fluorescence microscopy. *Nat. Methods* **22**, 1299–1310 (2025). doi:10.1038/s41592-025-02667-6

35. Zong, W. et al. Large-scale two-photon calcium imaging in freely moving mice. *Cell* **185**, 1240–1256.e30 (2022). doi:10.1016/j.cell.2022.02.017

36. Stringer, C. et al. Cellpose: a generalist algorithm for cellular segmentation. *Nat. Methods* **18**, 100–106 (2021). doi:10.1038/s41592-020-01018-x

37. Luo, Z. et al. Structured illumination-based super-resolution live-cell quantitative FRET imaging. *Photon. Res.* **11**, 887 (2023). doi:10.1364/prj.485521.

38. Sage, D. et al. Quantitative evaluation of software packages for single-molecule localization microscopy. *Nat. Methods* **12**, 717–724 (2015). doi:10.1038/nmeth.3442

39. Ovesný, M., Křížek, P., Borkovec, J., Švindrych, Z. & Hagen, G. M. ThunderSTORM: a comprehensive ImageJ plug-in for PALM and STORM data analysis and super-resolution imaging. *Bioinformatics* **30**, 2389–2390 (2014). doi:10.1093/bioinformatics/btu202

40. Rieger, B. & Stallinga, S. The Lateral and Axial Localization Uncertainty in Super-Resolution Light Microscopy. *ChemPhysChem* **15**, 664-670 (2014). doi:10.1002/cphc.201300711

41. Thevathasan, J. V. et al. Nuclear pores as versatile reference standards for quantitative superresolution microscopy. *Nat. Methods* **16**, 1045–1053 (2019). doi:10.1038/s41592-019-0574-9

42. Grußmayer, K., Lukes, T., Lasser, T. & Radenovic, A. Self-blinking dyes unlock high-order and multiplane super-resolution optical fluctuation imaging. *ACS Nano* **14**, 9156–9165 (2020). doi:10.1021/acsnano.0c04602

43. Tu, Z. et al. High-fidelity reconstruction in structured illumination microscopy through accurate noise modeling and multi-scale wavelet denoising. *Photon. Res.* **14**, 1383–1400 (2026). doi:10.1364/PRJ.585263.

## Methods

Open-source data

The ER and CCPs data in Fig. 1c were from the BioSR dataset[21] (https://doi.org/10.6084/m9.figshare.13264793). The mitochondria data in Fig. 1c were from BF-SIM[33]. The ER data in Extended Data Fig. 2 were from Dark sectioning[34]. The mitochondria data in Extended Data Fig. 4a and 4d were from UiT Open Research Data collection[44] (https://doi.org/10.18710/PDCLAS) and 3D-MP-SIM[45] respectively. The synthetic data imitating microtubules in Fig. 5b and 5c were from http://bigwww.epfl.ch/smlm/datasets/[38]. The FF data of microtubules in Fig. 6f were from self-blinking SOFI[42].

Open-source algorithms

Sparse deconvolution[17] (https://github.com/WeisongZhao/Sparse-SIM) was used for comparison in Fig. 1c and Extended Data Fig. 4. DPR[23] (https://github.com/biomicroscopy/DPR-Project) was used for comparison in Fig. 1c. MSSR[24] (https://github.com/MSSRSupport/MSSR) was used for comparison in Fig. 1c. Rolling ball in Fiji was used for background removal in Fig. 2k, Extended Data Fig. 4d and 6. Cellpose[36] (https://www.cellpose.org/) was used for detection of neurons in Extended Data Fig. 3. eSRRF[16] (https://github.com/HenriquesLab/NanoJ-eSRRF) was used for comparison in Fig. 5 and 6g. MLE fitting in ThunderSTORM[39] (https://code.google.com/p/thunder-storm/) was used for comparison in Fig. 5 and 6. Decorrelation analysis[31] (https://github.com/Ades91/ImDecorr.git) was used for resolution measurement in Fig. 7e,j and Extended Data Fig. 3d.

Cell culture and staining

For WF imaging of migrasome in Fig. 2a, before cell seeding, glass-bottom imaging chambers were pre-coated with fibronectin at 10 μg/ml. The coating was carried out at 37 °C for at least 30 minutes, with the duration not exceeding 2 hours. After coating, NRK cells stably overexpressing

Tspan4-GFP were seeded into the chambers. These cells were maintained in DMEM supplemented with 10% fetal bovine serum and 1× penicillin-streptomycin solution. The cultures were incubated at 37 °C in a 5% $CO_2$ atmosphere for approximately 15 hours, after which the samples were ready for subsequent imaging analysis.

For WF imaging of mitochondria in Fig. 2e and 2k, COS-7 cell line was purchased from Applied Biological Materials Inc. Cells were cultured in DMEM (Invitrogen, #11965-118) supplemented with 10% fetal bovine serum (FBS) (Gibco, #16010-159). To prevent bacterial contamination, 100 μg/ml penicillin and streptomycin (Invitrogen, #15140122) were added in the DMEM medium. Cells were grown under standard cell culture conditions (5% $CO_2$, humidified atmosphere at 37°C). For labeling, cells were seeded in a 35 mm cell dish with a glass bottom and incubated overnight. When the cells grown to a confluence of around 60%, cells were washed with PBS for three times and fresh culture medium containing PK Mito Orange (Nanjing Genvivo Biotech Co., Ltd.) at a concentration of around 200 nM was added into the cell dish. After additional incubation for 20 min, the COS-7 cells were washed with PBS for three times and fresh culture medium was added.

For FRET imaging in Fig. 3, EGFP (#74165) and mCherry (#176016) plasmids were obtained from Addgene. The mCherry-ActA plasmid was provided by David W. Andrews. GFP-ActA was generated by PCR amplification of the ActA coding sequence from mCherry-ActA and replacement of the Bak coding region in GFP-Bak. G17M-ActA was generated by replacing the stop codon in G17M with the ActA cloning sequence. U2OS cells were maintained in DMEM (Gibco) supplemented with 10% FBS and 1% gentamicin-amphotericin B at 37 °C and 5% CO2. Cells were plated in 20-mm glass-bottom dishes and transfected at 50–60% confluence using TurboFect (Fermentas) according to the manufacturer's instructions. Imaging was performed 24 h after transfection.

For 3D WF imaging of lipid droplets in Extended Data Fig. 6, live COS-7 cells were stained in DMEM+ containing 2 μM fluorescent probe Lipi-QA and 1% DMSO for 2 h in a $CO_2$ incubator. Then, the cells were washed three times with fresh medium to remove the free probe, and kept in HBSS for imaging.

Nuclear pore complex samples in Fig. 6a were prepared from U2OS cells stably expressing Nup96-SNAP (300444, Cell Line Services). Cells were fixed, permeabilized, and labeled with the SNAP-tag ligand BG-AF647 (S9136S, New England Biolabs; 1 μM) following established protocols[44]. Excess dye was removed by PBS washing, and samples were post-fixed prior to imaging.

## PEGASOS tissue clearing method

*Thy1-EGFP-M* Mouse brain samples in Extended Data Fig. 5 were collected and processed using PEGASOS tissue clearing kit (Leads Bio-Tech, http://www.leads-tech.com Catalogue number: PSK100N). A Mice were transcardially perfused with PBS followed by 4% paraformaldehyde (PFA). The brains were then dissected and fixed in 4% PFA overnight at 4 degrees. They were then processed following the kit protocol. Finally, samples were immersed in BB-PEG medium (solution 7 in the kit) for at least one day for clearing. Samples were immersed in the clearing medium for imaging.

## *In vivo* Two-photon imaging

Calcium imaging recordings in Extended Data Fig. 3 were obtained from the rerosplenial cortex (RSC) of mice using MINI2P (Thorlabs) during the course: "Imaging structure & function in the Nervous system" at Cold Spring Harbor Laboratory in 2024.

3D imaging of zebra finch in Fig. 4 was performed in accordance with the protocol approved by the government of Upper Bavaria (ROB-55.2-2532.Vet_02-18-142). Data were obtained from an adult male zebra finch (age: 115 days) from the breeding colony at the Max Planck institute for Biological Intelligence. Stereotaxic surgery was performed under anesthesia using isoflurane inhalation (0.8-1.8% at 0.5L $O_2$/min). scAAV-DJ/9-CMV-GFP (total volume: 600-800 nl) was injected into AreaX. A chronic cranial window with a headpiece was placed on the brain surface above HVC immediately after viral injection. Imaging was carried out with a two-photon microscope (MOM, Sutter instruments) coupled to a pulsed Ti:Sapphire laser (Mai Tai DeepSee, Spectra-Physics) and controlled by ScanImage 5.4 at 1024×1024 pixels size. A constant average power (56 mW) was applied on the brain tissue. Images were acquired using a 25x/NA:0.95 water immersion objective lens (Leica) and photomultiplier tubes (Hamamatsu). Time lapse images were acquired when animal was anesthetized using isoflurane inhalation and the head was fixed under the two-photon microscope objective. Animal was kept warm on a constant temperature pad (160x160 mm, 12V/6W, Thermo) and wrapped in a thin gauze blanket.

WF, SIM and confocal imaging

The commercial Nikon ECLIPSE T*i*2-U microscope equipped with a sCMOS camera (ORCA-Flash4.0 V3, 100 fps @ 2,048 × 2,048 pixels, 16 bits, Hamamatsu, Japan) is used for WF imaging. The SR-SIM imaging is conducted on our previous custom-built SIM system[47]. The confocal imaging is conducted on a commercial confocal laser scanning microscope (Leica SP8 DIVE).

SMLM imaging

SMLM imaging in Fig. 6a was performed on a custom-built single-molecule localization microscopy system as described before[48]. Fluorophores were driven into a stochastic blinking regime using a 640 nm laser (iBEAM-SMART-640-S-HP, 200 mW, TOPTICA Photonics), with a 405 nm laser (iBEAM-SMART-405-S, 150 mW, TOPTICA Photonics) used to control the activation density. A 785 nm near-infrared laser (iBEAM-SMART-785-S, TOPTICA Photonics) was introduced for focus stabilization. Single-molecule fluorescence was collected using a high numerical aperture oil immersion objective (NA 1.5, UPLSAPO 100XOHR, Olympus) and recorded with an sCMOS camera (ORCA-Flash4.0 V3, HAMAMATSU). Samples were imaged in an oxygen-scavenging buffer containing glucose oxidase (G7141, Sigma), catalase (C100, Sigma), and cysteamine (30070, Sigma) to promote photoswitching. 100,000 frames were recorded at an exposure time of 15 ms per frame.

## Quantification of FRET signals and system calibration

For three-channel FRET quantification in Fig. 3, background was subtracted from each image using the mean intensity of a cell-free region, and a binary mask was applied to exclude background pixels[37]. The donor-centric FRET efficiency ($E_D$) and the concentration ratio of total acceptor to donor ($R_C$) were calculated according to Eqs. (6) and (7) respectively.

$$E_D = \frac{F_C}{F_C + G \cdot I_{DD}}, \tag{6}$$

$$R_C = \frac{k \cdot I_{AA}}{F_C / G + I_{DD}}, \tag{7}$$

The corrected acceptor-sensitized emission ($F_C$) was calculated according to Eq. (8) from the acceptor-channel signal acquired under donor excitation after correction for spectral crosstalk.

$$F_C = I_{DA} - a(I_{AA} - c \cdot I_{DD}) - d(I_{DD} - b \cdot I_{AA}), \tag{8}$$

In these calculations, $G$ denotes the sensitivity quenching factor, $k$ denotes the concentration correction factor, $I_{DD}$ denotes the donor-channel intensity under donor excitation, and $I_{AA}$ denotes the acceptor-channel intensity under acceptor excitation. The spectral crosstalk coefficients $a$, $b$, $c$, *d*. were determined experimentally using donor-only and acceptor-only samples.

Spectral crosstalk coefficients (*a*, *b*, *c*, *d*.) were measured in live U2OS cells expressing GFP or mCherry alone. The calibration factors $G$ and $k$ were determined using the mPb-G[49] method with G17M and G32M. Analysis of at least 20 live U2OS cells yielded : $a = 0.041$, $b = 0.001$, $c = 0.002$, $d = 0.055$, $G = 0.382$, and $k = 1.957$.

Evaluation of image quality

We used PSNR to evaluate the similarity to GT image:

$$PSNR(\text{dB}) = 10\log_{10}\frac{1}{(I-I_{GT})^2}, \tag{9}$$

The SNR of image was calculated as:

$$SNR(\text{dB}) = 10\log_{10}\frac{std[I]^2}{stdfilt[I]^2}, \tag{10}$$

The fidelity of the reconstructed image was measured by Fourier correlation within diffraction limit with the raw image:

$$\text{Fidelity} = \frac{\sum_{\|r\|<f_c} F(r) \cdot F_{raw}(r)^*}{\sqrt{\sum_{\|r\|<f_c} F^2(r) \cdot \sum_{\|r\|<f_c} F_{raw}^2(r)}}, \tag{11}$$

where $f_c$ is the cutoff frequency, $F$ and $F_{raw}$ are the Fourier transforms of the reconstructed image and the raw image. A fidelity value closer to 1 indicates better fidelity of the reconstructed image.

**Data Availability**

The imaging data of mouse brain neurons in Extended Data Fig. 5 is available at https://doi.org/10.6084/m9.figshare.32071824. The imaging data of mitochondria in Fig. 2e and some open-source data are available at https://github.com/M-R-Wang/SoLo as example data. Other datasets used and analyzed during the current study are available from the corresponding author upon reasonable request.

**Code Availability**

The softwares for SoLo and NL-SoLo are available at https://github.com/M-R-Wang/SoLo.

**References**

44. Opstad, I. S. et al. Three-dimensional structured illumination microscopy data of mitochondria and lysosomes in cardiomyoblasts under normal and galactose-adapted conditions. *Sci. Data* **9**, 98 (2022). doi:10.1038/s41597-022-01207-7

45. Chen, Q. et al. Fast, three-dimensional, live-cell super-resolution imaging with multiplane structured illumination microscopy. *Nat. Photon.* **19**, 567–576 (2025). doi:10.1038/s41566-025-01638-9

46. Fu, S. et al. Field-dependent deep learning enables high-throughput whole-cell 3D super-resolution imaging. *Nat. Methods* **20**, 459–468 (2023). doi:10.1038/s41592-023-01775-5

47. Wang, Z. et al. High-speed image reconstruction for optically sectioned, super-resolution structured illumination microscopy. *Adv. Photon.* **4**, 026003 (2022). doi:10.1117/1.AP.4.2.026003

48. Fu, S. et al. Deformable mirror-based optimal PSF engineering for 3D super-resolution imaging. *Opt. Lett.* **47**, 3031 (2022). doi:10.1364/OL.460949

49. Zhang, J. et al. Reliable measurement of the FRET sensitized-quenching transition factor for FRET quantification in living cells. *Micron* **88**, 7–15 (2016). doi:10.1016/j.micron.2016.04.005.

## Acknowledgements

This work was supported by the National Key Research and Development Program of China (2023YFF0722600 and 2022YFF0712500), the Natural Science Foundation of China (62135003, 62435007, 62205267, and 62205265), and the Natural Science Basic Research Program of Shanxi (2024JC-YBMS494 and 2022JM-321). We would like to thank the 2024 CSH Imaging Structure and Function in the Nervous System Course for providing us with the MINI2P RSC mouse data. We thank D. Hou and J. Ma for sharing the public dataset of self-blinking SOFI.

## Author Contributions

M.W. and M.L. conceived the project. M.W. developed the algorithm and implemented the software. M.W. conducted the imaging experiments of mitochondria and mouse barin. S.M. conducted the 2P imaging experiment. Z.L. conducted the FRET imaging experiment and analysis. W.S. and Y.L. conducted the SMLM imaging experiment. C.L. conducted the imaging experiments of migrasome and lipid droplets. M.W. performed the data analysis and wrote the paper. M.L. supervised the research project. All authors participated in reading and editing of the final paper.

## Competing interests

M.L. and M.W. are applying for a patent on the SoLo algorithm. All other authors declare they have no competing interests.

# Figures and Tables

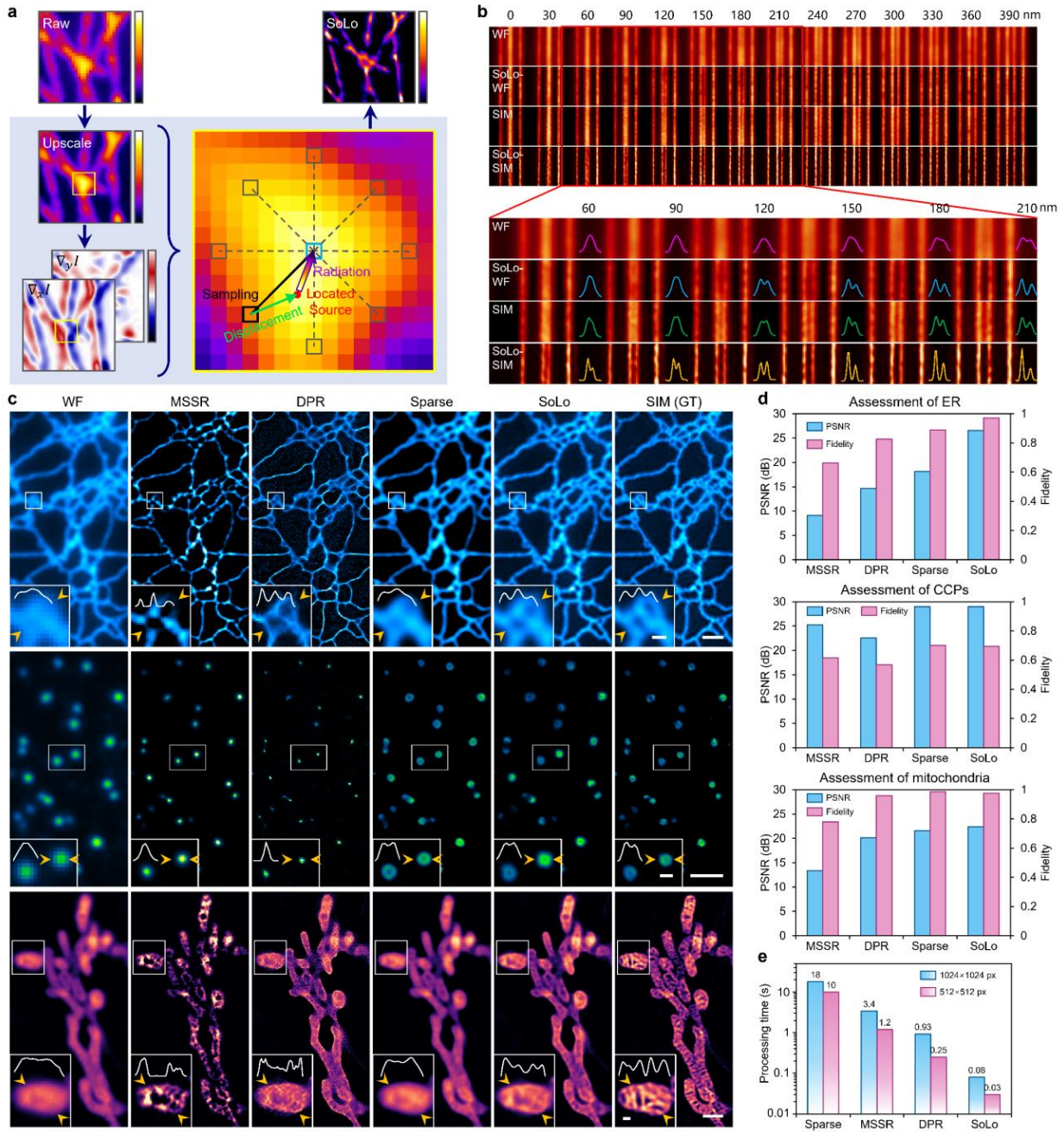

**Figure 1. Principle and performance of SoLo. a**, Workflow diagram of the SoLo algorithm using a WF image of microtubule as an example. The process begins with upscaling the raw image and computing its gradient fields. For each detection position (blue box), the algorithm samples 8

peripheral points (black boxes), calculates their displacement vectors to localize potential source positions (red point), then applies distance-attenuating radiation back to the detection location. This operation is iteratively applied across the entire image to generate the final super-resolution SoLo reconstruction. **b**, Resolution of fluorescent line pairs with varying spacings. SoLo was applied to both WF and SIM imaging, enhancing their resolutions to 120nm and 60nm respectively. **c**, Performance comparison between MSSR, DPR, Sparse deconvolution, and SoLo on WF images of ER, CCPs, and mitochondria. The SR-SIM images serve as GT. **d**, Quantified assessment of each algorithm including PSNR with SIM and fidelity with WF. **e**, Comparison of the processing time between each algorithm on images of 512×512 pixels and 1024×1024 pixels respectively. Scale bar, 1 μm (**c**), 200 nm (magnified regions in **c**).

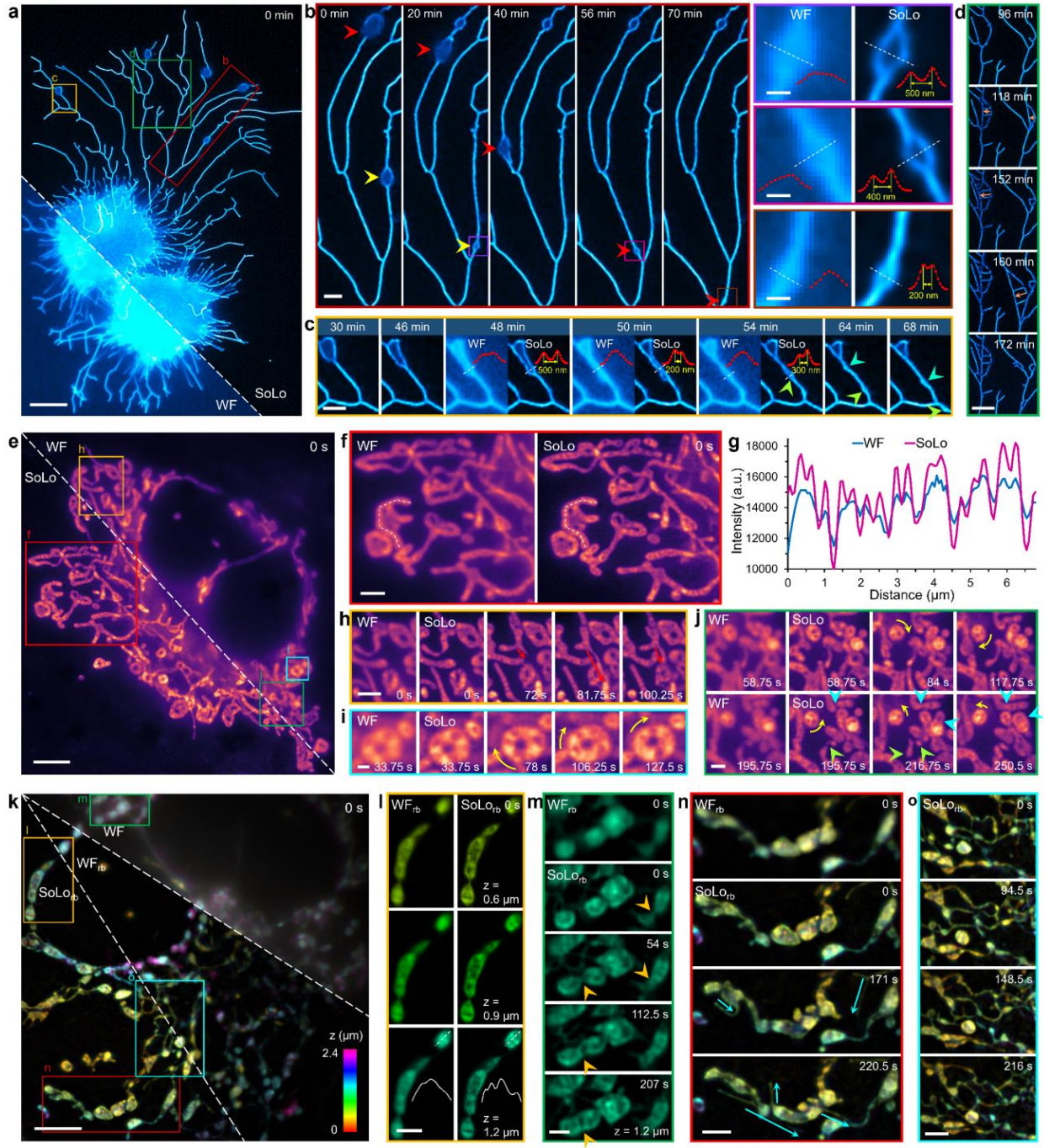

**Figure 2. SoLo achieves single-frame super-resolution in live-cell WF imaging. a–d,** Long-term recording of migrasome retrieval in Tspan4-GFP NRK cells. WF and SoLo-enhanced images are compared, along with intensity profiles along the dashed lines. Arrows indicate the process of migrasome transport along the retraction fibers toward the cell body (**b, c**), as well as changes in the retraction fibers (**d**). **e-j**, Comparison of WF imaging and real-time SR results by SoLo of mitochondrial dynamics in COS-7 cells stained with PK Mito Orange. The intensity profile along the mitochondria in **f** is plotted in **g** to illustrate the enhancement of cristae structures by SoLo. Enlarged images at multiple time points show mitochondrial extension and kissing (**h**), rotation (**i**, **j**), as well as fission and fusion (**j**). **k**, Height-coded colored MIP of multi-layer scanning WF imaging of mitochondrial dynamics in COS-7 cells stained with PK Mito Orange, where $WF_{rb}$ and $SoLo_{rb}$ represent background-removed images using the rolling ball method. **l**, Comparison of $WF_{rb}$ and $SoLo_{rb}$ of mitochondria at different z-slices from **k**. **m**, Fusion process of mitochondria in the z = 1.2 μm layer from **k**. **n**, Extension behavior of mitochondria in **k**. **o**, Temporal changes of the mitochondrial network in **k**. Scale bar, 10 μm (**a**), 5 μm (**d, e, k**), 2 μm (**b, c, f, l, n, o**), 1 μm (**j, m**), 500 nm (magnified region in **b, i**).

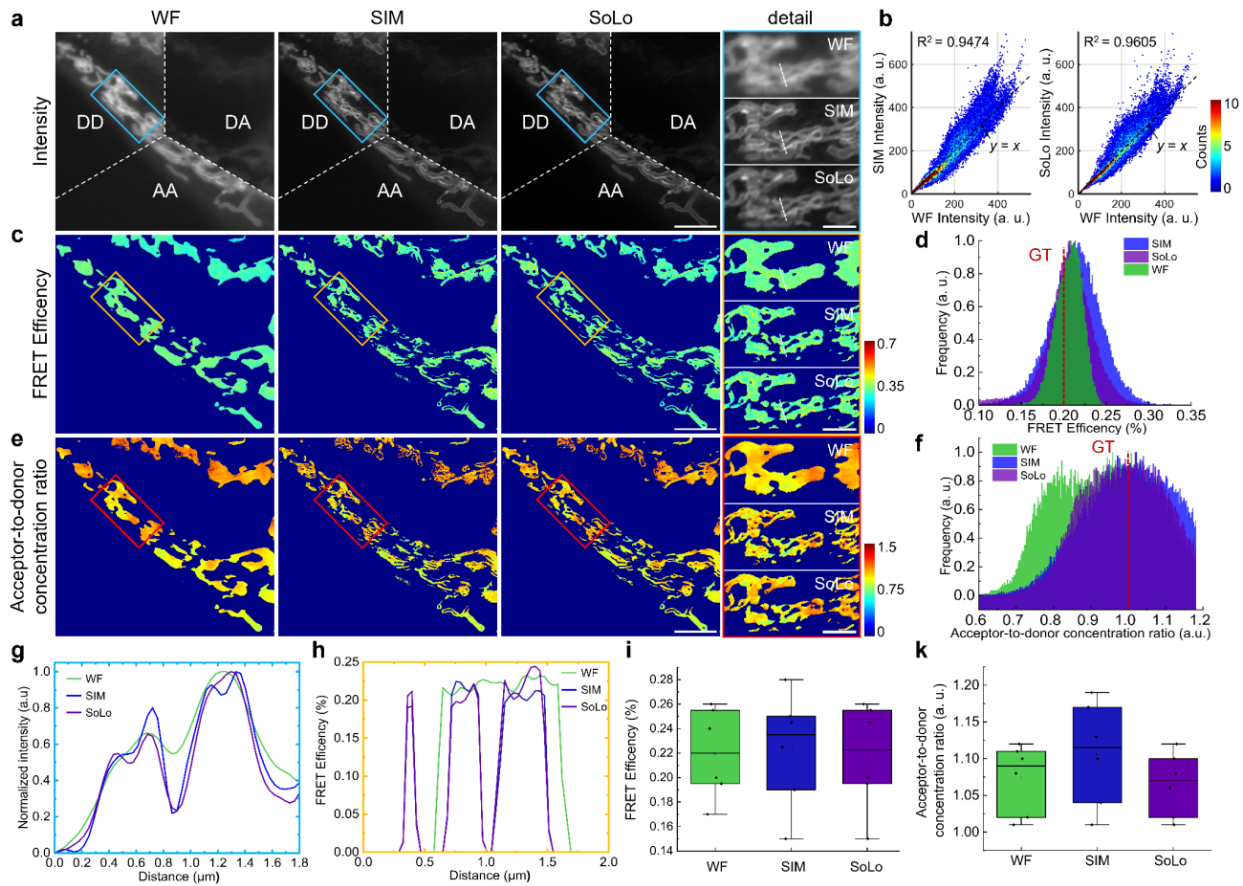

**Figure 3. Comparative evaluation of SoLo and SIM in quantitative super-resolution FRET imaging. a**, Multichannel intensity images (DD, DA, AA) of ACTA-G17M in live U2OS cells obtained via WF, SIM, and SoLo. **b**, Density scatter plot of the correlation between WF versus SIM and SoLo intensities of DD channel. **c**, Sub-diffraction spatial maps of FRET efficiency $E_D$. **d**, Histograms of $E_D$ distributions. **e**, Sub-diffraction spatial maps of concentration ratio $R_C$. **f**, Histograms of $R_C$ distributions. **g-h**, Line profiles of normalized intensity and FRET efficiency across mitochondrial structures corresponding to the regions in **a** and **c**. **i-k**, Statistical distribution of mean $E_D$ and $R_C$ values across multiple fields of view (n = 7). Scale bar, 5 μm, 2μm (detail).

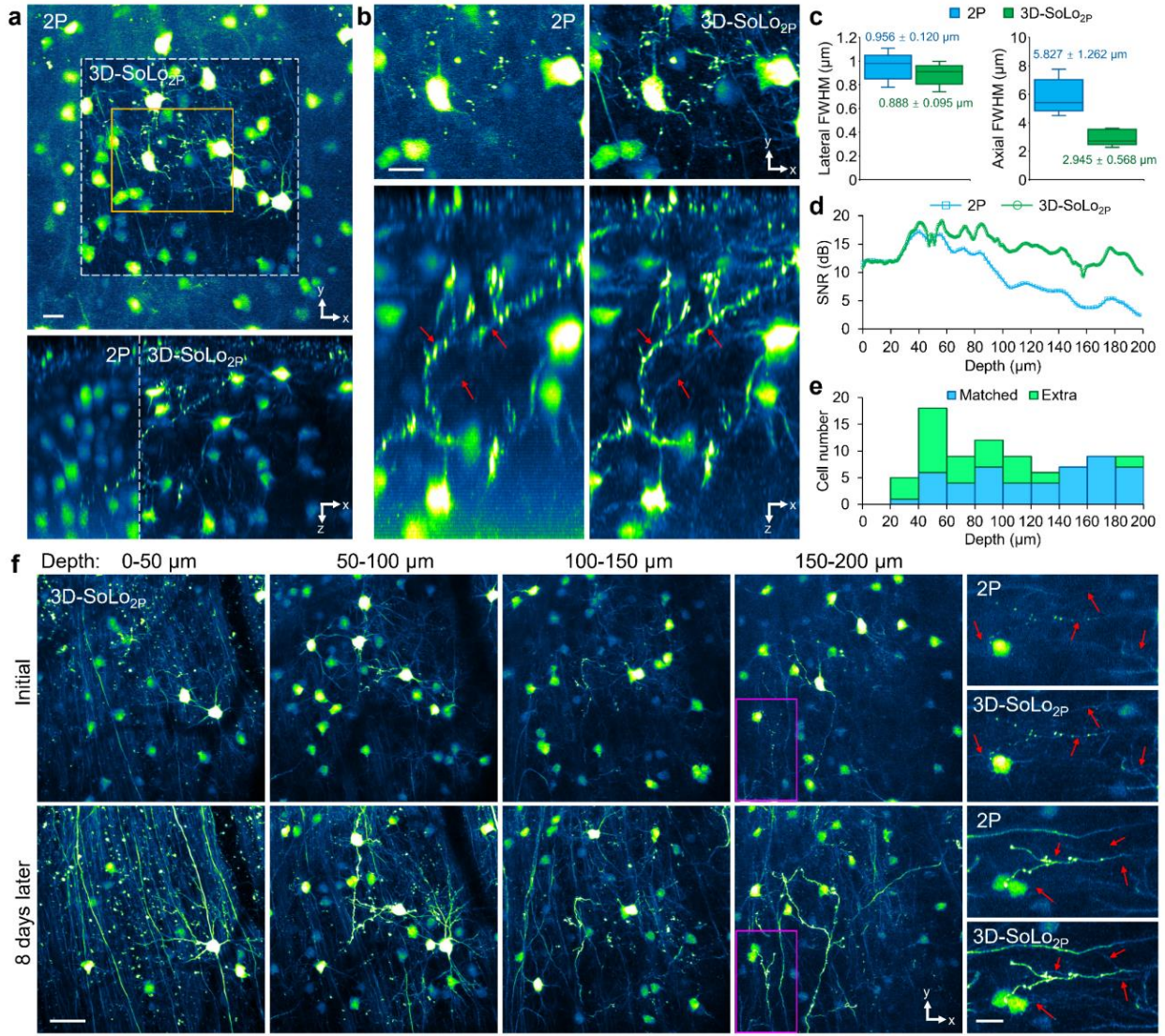

**Figure 4. 3D resolution improvement in 2P imaging of zebra finch brain neurons using 3D-SoLo. a**, Maximum intensity projection (MIP) images along the z-axis and y-axis for 2P and 3D-SoLo processed stack. **b**, Comparison of MIP images along the z-axis and y-axis for the yellow-boxed region in **a** between 2P and 3D-SoLo$_{2P}$. **c**, Comparison of measured lateral and axial FWHM of the PSF for 2P and 3D-SoLo$_{2P}$. Measurements were performed at multiple locations of fine nerve fibers. **d**, SNR of 2P and 3D-SoLo$_{2P}$ images at varying imaging depths. **e**, Histogram of the neuron number detected at different depths. Blue bars represent neurons detected by both 2P and 3D-SoLo$_{2P}$ (matched), and green bars represent neurons additionally detected after using 3D-SoLo

(extra). **f**, MIP images comparing neuronal morphology across different depth ranges after 8 days of imaging. Scale bar, 20 μm (**a**, **b**, magnified regions in **f**), 50 μm (**f**). Red arrows highlight morphological features that were enhanced by 3D-SoLo$_{2P}$, revealing finer neuronal structures previously obscured by noise.

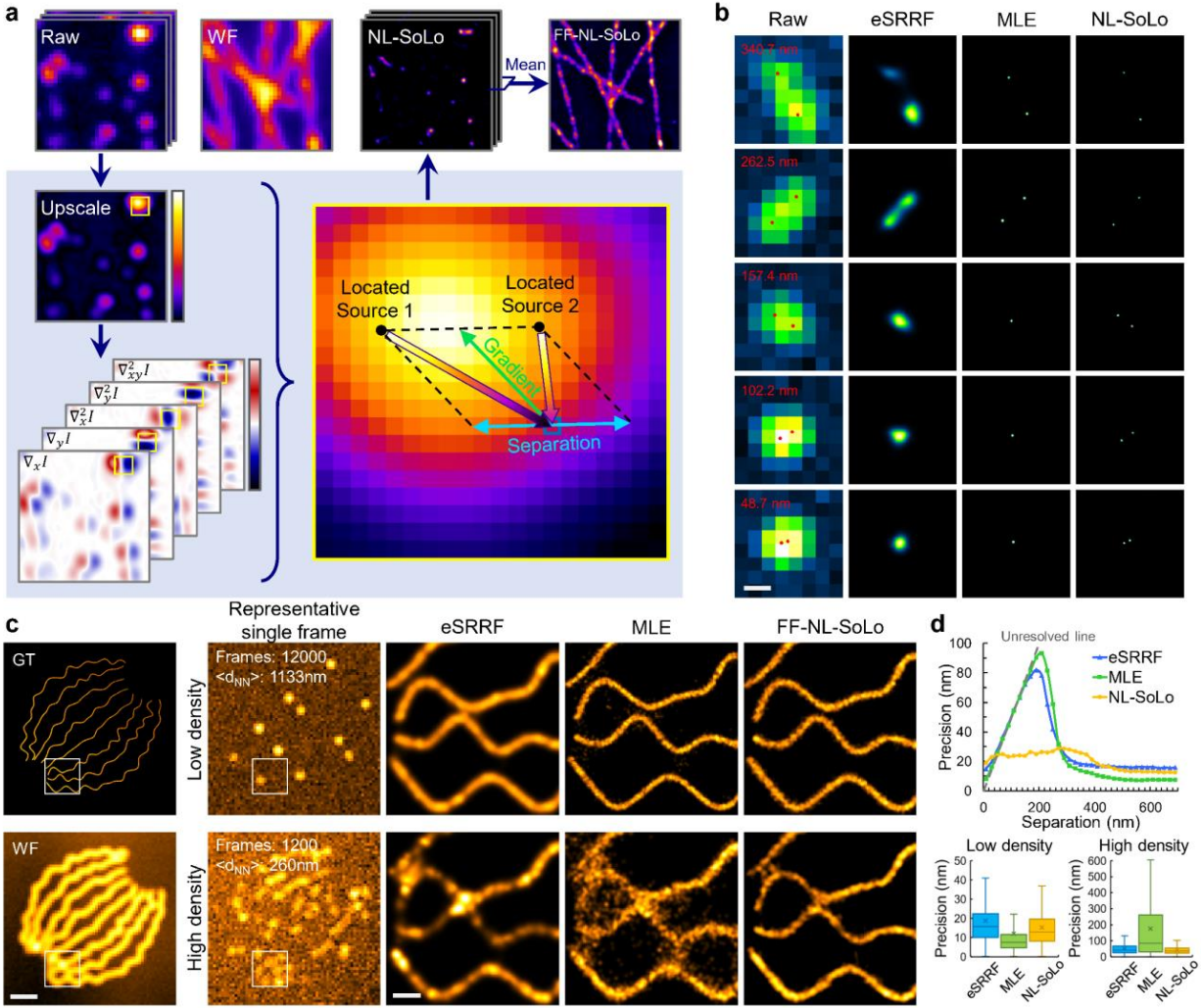

**Figure 5. Enhanced super-resolution capability of NL-SoLo applied to fluorescence fluctuation imaging. a**, Workflow diagram of the NL-SoLo algorithm using fluorescence fluctuation (FF) images of microtubules as an example. For each raw image, it is first upscaled, and its first- and second-order gradient fields are computed. The gradient and separation vectors at each pixel locate two potential source positions, with distance-attenuating radiation back to the detection location to generate the NL-SoLo image. Averaging all NL-SoLo images produces the final FF-NL-SoLo super-resolution image. **b**, Comparison of different methods (eSRRF, multi-emitter maximum likelihood estimation (MLE) fitting, and NL-SoLo) on resolving pairs of fluorescent molecules with varying inter-particle distances. The raw images are single-frame simulations, with red dots marking the GT positions of fluorescent molecules. **c**, Comparison of different methods on super-resolution reconstruction for simulated FF imaging of microtubule. Reconstructions were performed separately for low and high fluorophore density conditions, and the result details within the white boxed region are displayed. **d**, Localization precision of each method. The top panel plots the relationship between localization precision and nearest-neighbor separation for each method, while the bottom panel illustrates the distinct precision characteristics of the methods under low and high density conditions. Scale bar, 1 μm (**c**), 200 nm (**b**, magnified regions in **c**).

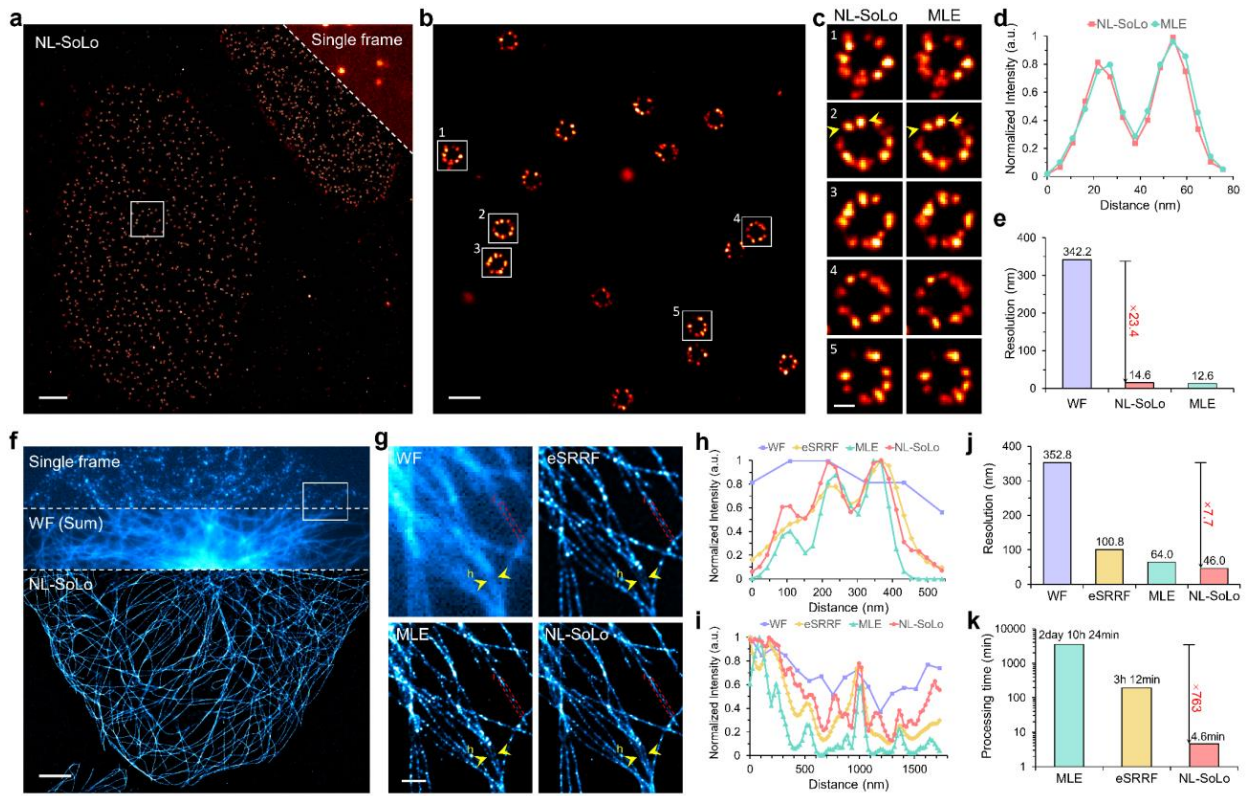

**Figure 6. High-precision FF reconstruction achieved by NL-SoLo under both low- and high-density excitation. a**, Raw single frame of Nup96 under low-density excitation, and the FF image reconstructed from 100,000 frames using NL-SoLo. **b**, Magnified image of the white boxed area in **a**. **c**, Comparison of NL-SoLo and MLE reconstruction results for the five NPCs in **b**. **d**, Intensity distribution between the two corners of the second NPC in **c**. **e**, Comparison of resolution obtained by decorrelation analysis of the raw WF image, NL-SoLo and MLE reconstructed images of Nup96. **f**, Raw single frame of microtubules under high-density excitation, the WF image obtained by sum of the frames, and the FF image reconstructed from 8,000 frames using NL-SoLo. **g**, Comparison of WF image and reconstruction results from eSRRF, MLE, and NL-SoLo of the white boxed area in **f**. **h**, Comparison of intensity distributions along the yellow arrow in **g** among different methods. **i**, Comparison of intensity distributions along the microtubule indicated by the red box in **g** among different methods. **j**, Comparison of resolution obtained by decorrelation

analysis of the microtubule images reconstructed by different methods. **k**, Comparison of processing times among different methods, where the raw images are 500×500 pixels, and both eSRRF and NL-SoLo are upsampled by a factor of 5, tested on an RTX3070 GPU. Scale bar, 2 μm (**a**), 200 nm (**b**), 50 nm (**c**), 5 μm (**f**), 1 μm (**g**).

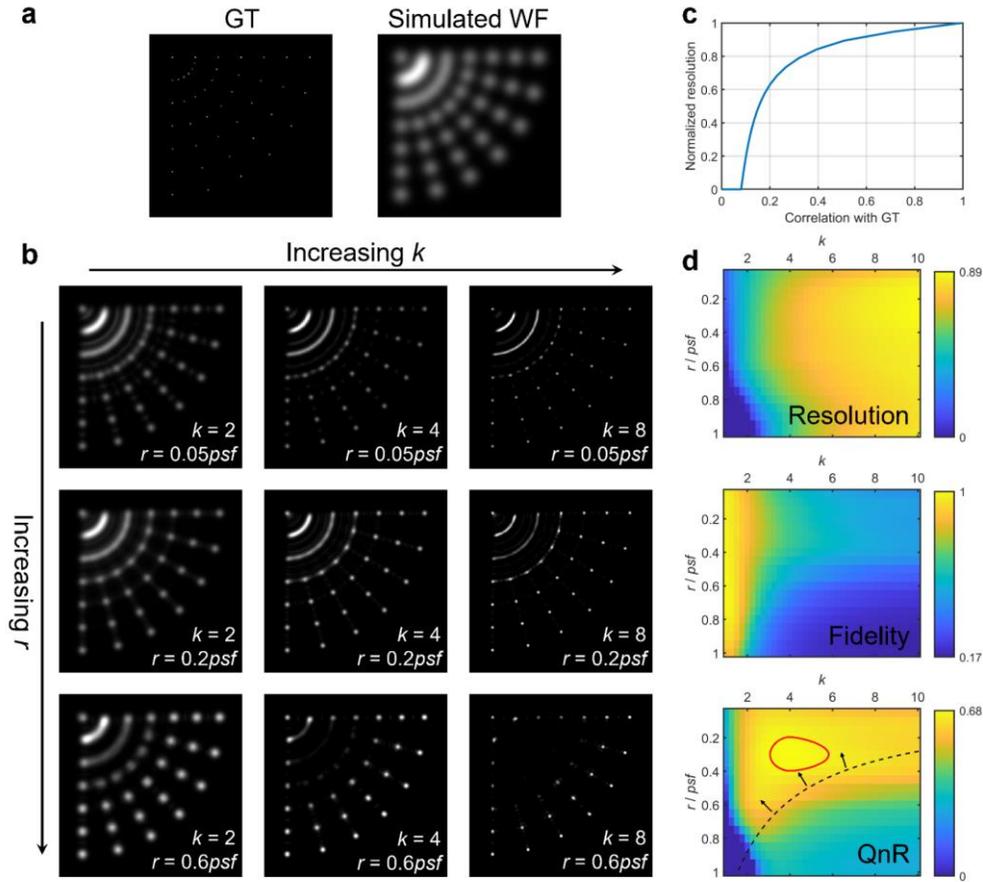

**Extended Data Figure 1. Optimal parameters determination of SoLo through dot-array simulation. a**, Ground truth (GT) image of the dot array and its simulated wide-field (WF) image. **b**, Effects of varying $r$ and $k$ on the output image. **c**, Relationship establishment between normalized resolution and correlation with GT. **d**, Resolution, fidelity, and QnR map with regard to $r$ and $k$. The red contour encloses the parameter range where QnR approaches its maximum value, and the dashed line indicates the curve $r \cdot k = 2.6psf$.

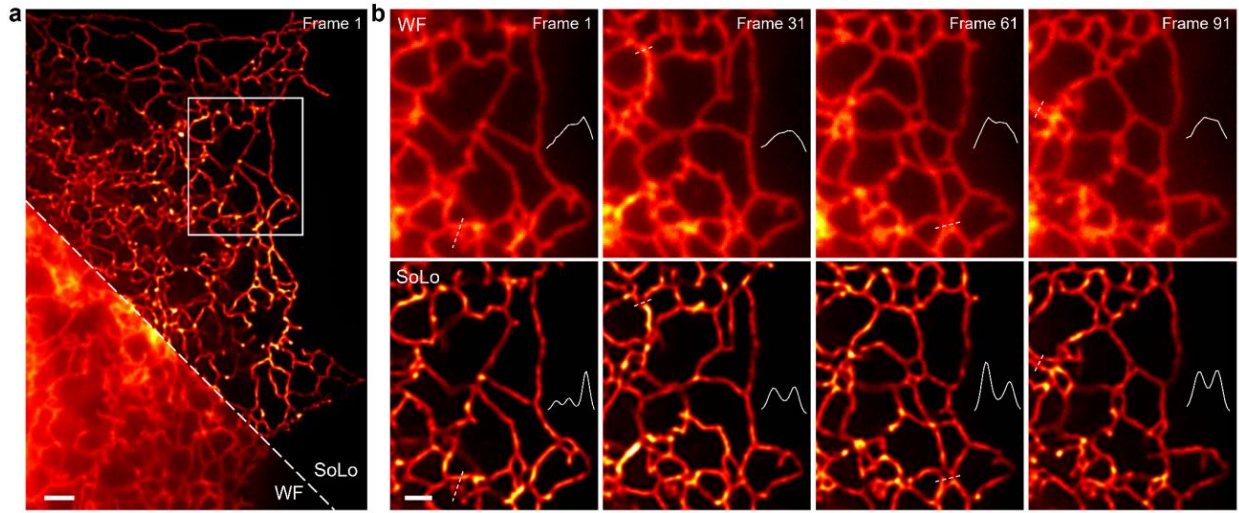

**Extended Data Figure 2. Comparison of WF and SoLo of ER tubes in live COS-7 cells with a plasmid expressing GFP-tagged KDEL protein. a**, The WF and SoLo images of the ER at the initial time point. **b**, WF and SoLo comparison of the boxed region in **a** and ER dynamics at different time points, with the intensity profile along the dashed line. Scale bar, 2 μm (**a**), 1 μm (**b**).

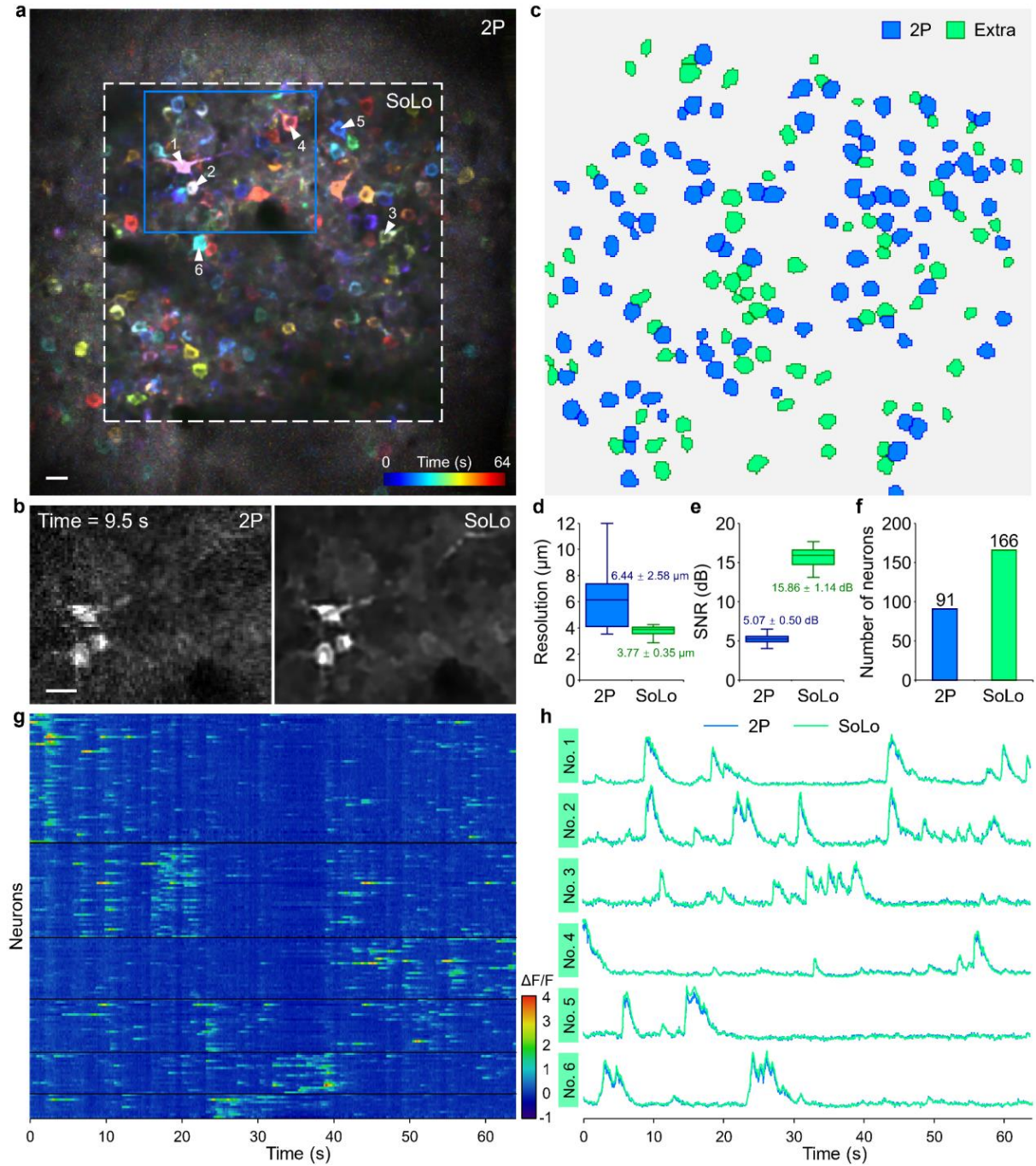

**Extended Data Figure 3. SoLo enchances calcium imaging in freely moving mouse. a**, Comparision of raw and SoLo-enhanced (within the dashed-outlined region) images reveals improved signal fidelity. **b**, Zoom-in of the blue-boxed region in **a** at 9.5 s, highlighting enhanced structural detail after SoLo processing. **c**, SoLo-enhancement enables detection of additional

neurons using Cellpose. **d**, Comparison of resolution measured by decorrelation analysis between 2P and SoLo-enhanced images. **e**, Comparison of SNR between 2P and SoLo-enhanced images. **f**, Comparison of the number of neurons detected in 2P and SoLo-enhanced images. **g**, Heatmap of calcium signals from all neurons detected by SoLo, clustered using k-means. ΔF/F represents the relative change in fluorescence, with the median value used as the baseline. **h**, Traces of calcium transients from six neurons marked in **a**. Scale bar, 20 μm (**a, b**).

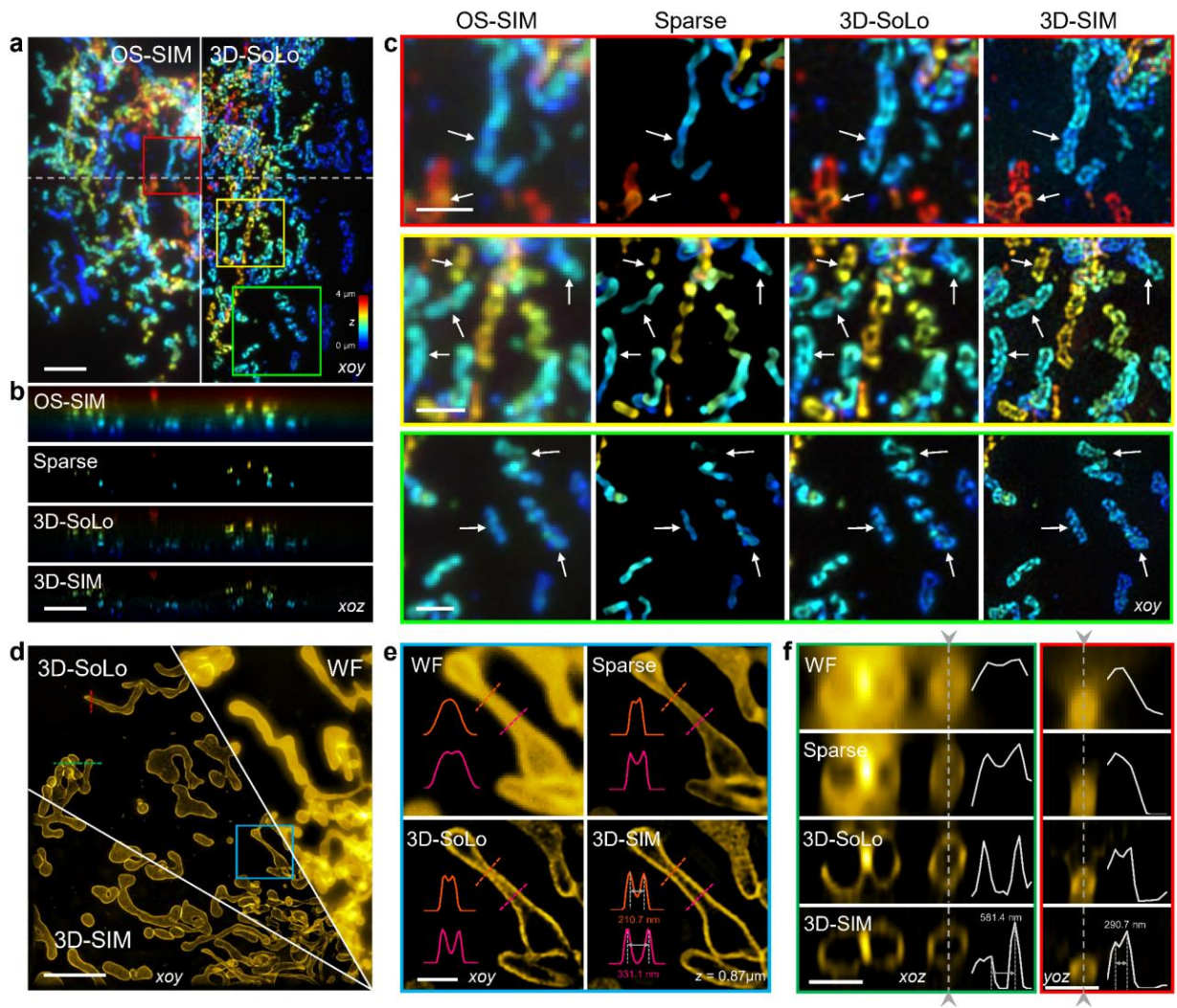

**Extended Data Figure 4. Comparison of 3D-SoLo and 3D-SIM for mitochondrial outer membrane imaging. a**, MIP images of the mitochondrial outer membrane in H9c2 cells acquired by OS-SIM and after 3D-SoLo processing, with color coded according to height. **b**, Comparison of *xoz* cross-sections along the dashed line in **a** for OS-SIM, Sparse, 3D-SoLo, and 3D-SIM. **c**, Comparison of height color-coded MIP images for OS-SIM, Sparse, 3D-SoLo, and 3D-SIM in the three boxed regions in **a**. **d**, MIP images of the mitochondrial outer membrane in COS-7 cells acquired by WF, after 3D-SoLo processing, and by 3D-SIM, with an imaging depth of 1.55 μm. **e**, Comparison of single-layer images for WF, Sparse, 3D-SoLo, and 3D-SIM at $z = 0.87$ μm in the boxed region in **d**. **f**, Comparison of *xoz* and *yoz* cross-sections along the dashed lines in **d** for WF, Sparse, 3D-SoLo, and 3D-SIM. Scale bar, 5 μm (**a**, **b**, **d**), 2 μm (**c**), 1 μm (**e**, **f**).

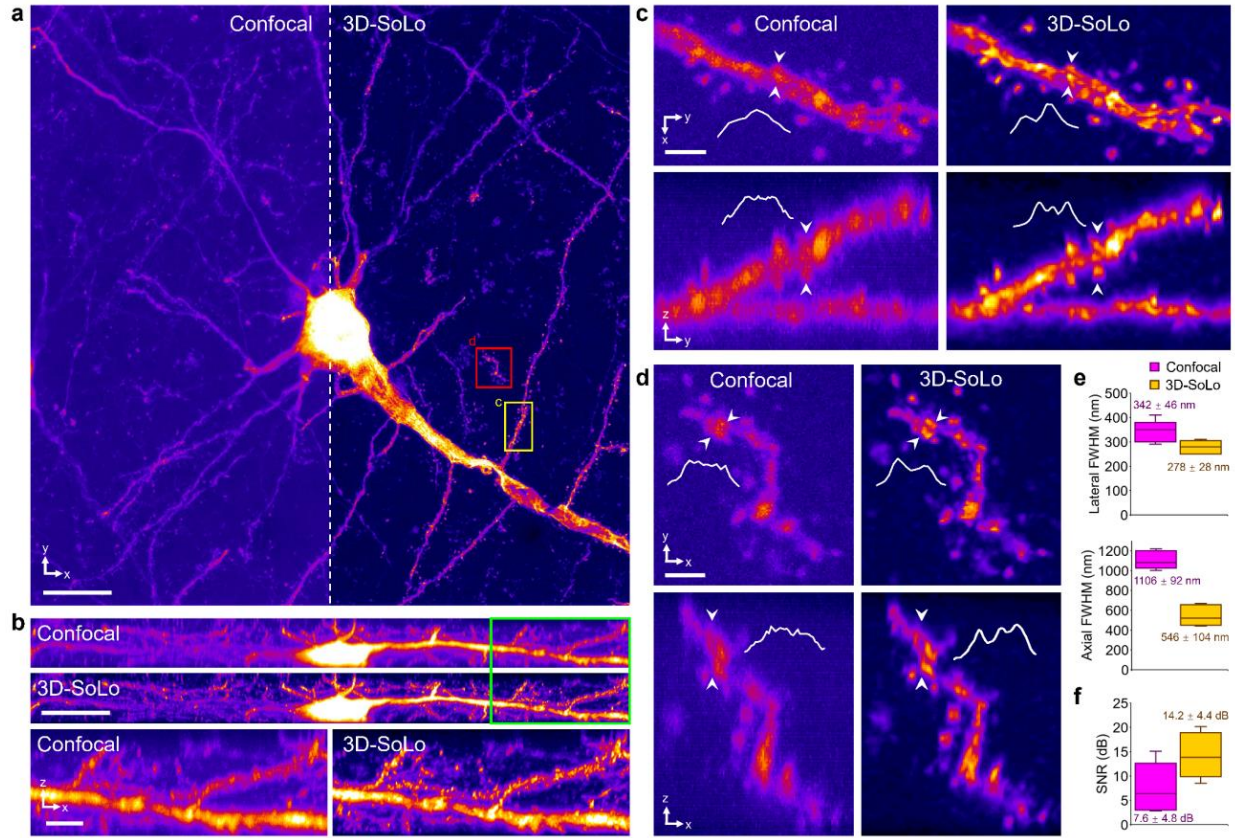

**Extended Data Figure 5. 3D-SoLo enhancement of confocal imaging of mouse brain neurons. a,** Z-axis MIP images of the confocal and 3D-SoLo enhanced result. **b,** Comparison of y-axis MIP images between confocal and 3D-SoLo, along with magnified views of the green boxed region. **c-d**, Comparison between confocal and 3D-SoLo for the z-axis and x/y-axis MIP images corresponding to the boxed regions in **a**. **e**, Comparison of measured lateral and axial FWHM of the PSF for confocal and 3D-SoLo. Measurements were performed at multiple locations of fine nerve fibers. **f,** Comparison of the SNR of confocal and 3D-SoLo. Scale bar, 20 μm (**a**, **b**), 5 μm (magnified region in **b**), 2 μm (**c**, **d**)

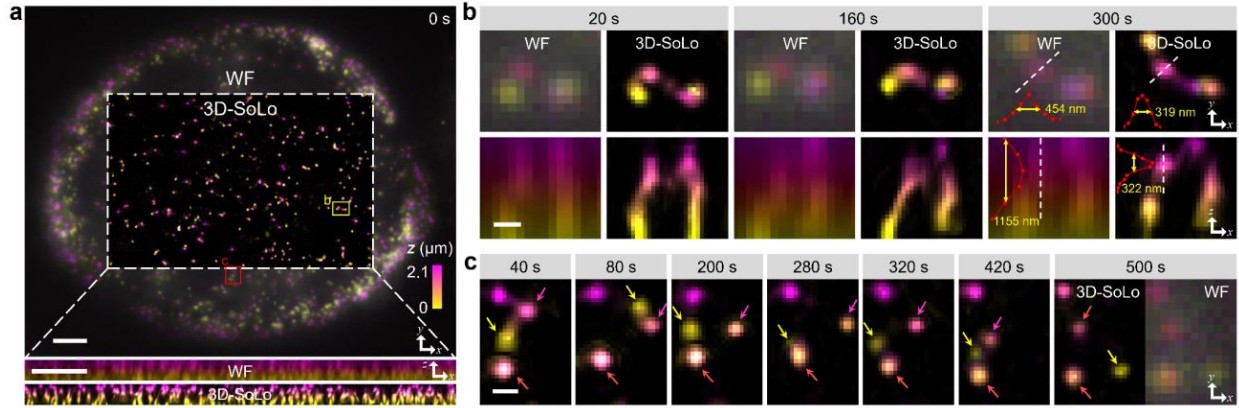

**Extended Data Figure 6. Comparison of 3D WF imaging and 3D-SoLo-enhanced results of lipid droplets in live COS-7 cells. a**, Comparison of WF and 3D-SoLo for their z-axis MIPs and y-axis MIPs of the dashed boxed region with height-coded color. **b**, Comparison of z-axis and y-axis MIPs and intensity profiles along the dashed line between WF and 3D-SoLo in the yellow boxed region in **a** at different time points. **c**, The movement of lipid droplets over time in the red boxed region in **a**. Scale bar, 5 μm (**a**), 500 nm (**b, c**)